\documentclass[preprint,floats,aps,showpacs,epsf]{revtex4}
\usepackage{epsfig}

\def\be{\begin{equation}}
\def\ee{\end{equation}}
\def\bdm{\begin{displaymath}}
\def\edm{\end{displaymath}}
\def\bea{\begin{eqnarray}}
\def\eea{\end{eqnarray}}
\def\nn{\nonumber}

\def\s{\sigma}

\newcommand{\p}{\partial}

\newcommand{\la}{\langle}
\newcommand{\ra}{\rangle}
\newcommand{\rd}{\mbox{d}}
\newcommand{\ri}{\mbox{i}}
\newcommand{\re}{\mbox{e}}

\newcommand{\bN}{{\bf N}}

\begin{document}
\draft
\title{Spinons in more than one dimension:  Resonance  Valence Bond state stabilized 
by frustration}
\author{ A. A. Nersesyan$^1$ and A. M. Tsvelik$^2$ }
\affiliation{$^1$ The Abdus Salam International Centre for 
Theoretical Physics, Strada Costiera 11, 34100 Trieste, Italy\\
$^2$ Department of  Physics, Brookhaven 
National Laboratory, Upton, NY 11973-5000, USA}
\date{\today}
\begin{abstract}
For two spatially  anisotropic, SU(2)-invariant models of frustrated  magnets 
in arbitrary space dimension  we present a non-perturbative proof of 
the existence of 
neutral spin-1/2 excitations (spinons).  In one model the frustration 
is static and based 
on fine tuning of the coupling constants, whereas in the other  
it is dynamic and does not require adjusting of the model parameters.  
For both models we derive a low-energy  effective action which does 
not contain any constraints. Though our models admit the standard gauge 
theory treatment, we  follow  an alternative approach based on Abelian 
and non-Abelian bosonization. We prove the existence of propagating 
spin-1/2 excitations (spinons) and consider in detail certain exactly 
solvable limits. A qualitative discussion of the 
most general case is also presented. 
\end{abstract}
\maketitle

\section{Introduction}

 Excitations with fractional quantum numbers in general and  neutral excitations with 
spin--1/2 in particular constitute an essential ingredient of many theoretical 
approaches aimed
to explain the non-Fermi liquid behavior of   
quasi-two-dimensional copper oxide materials. 
These approaches try 
to generalize for higher dimensions  
such mechanisms  as spin-charge separation and quantum number fractionalization, 
the phenomena well understood and adequately described
for one-dimensional systems (see recent review articles 
\cite{kivelson1}, \cite{kivelson+}, \cite{wilczek} and references therein).

As a part of the general program to construct a theory of copper oxides,  there 
have been many 
attempts to find neutral spin--1/2 excitations (spinons) in purely magnetic systems, 
that is to find higher-dimensional 
analogues of 1D spin S=1/2 Heisenberg antiferromagnet. Such 
generalization was  qualitatively outlined by Anderson \cite{anderson} in the form of  
the famous 
Resonant Valence Bond (RVB) state. This is a {\it spin-liquid} state which  
breaks neither translational nor spin rotational symmetry. Building on this proposal, 
Kivelson, Rokhsar and Sethna showed \cite{SRK} that such state,
if exists, must support neutral spin--1/2 excitations.

Despite more than  a decade of strong efforts, no realization of an RVB state 
supporting excitations with fractional spin 
in  D $>$ 1 Heisenberg magnets with short-range exchange interactions has been 
found in  models with interactions of the {\it Heisenberg} type. We exclude from 
the consideration  models of {\it quantum dimers}, where such excitations 
have been shown to exist \cite{sondhi},  and other recently suggested models with 
fractionalized excitations, which are not microscopic electronic models (see, 
for example \cite{fisher},\cite{nayack}), as well as
 generalizations for symmetries higher than SU(2). We also do not consider magnets with 
{\it incommensurate} ordering (such as a spiral one) which also support spin-1/2 
excitations \cite{chubukov1},\cite{chubukov2}.  Strong qualitative arguments against the
RVB scenario have been presented by Read and 
Sachdev \cite{ReadSachdev}, who argued that the most likely mechanism  for disordering 
an  antiferromagnetic state is the {\it spin-Peierls} one. Here, although spin 
rotational symmetry 
is preserved, translational symmetry 
is broken since the spin system 
undergoes an explicit dimerization at a finite temperature. 
That makes the existence of deconfined spinons in the low-temperature phase impossible.

 An interesting by-product of the RVB scenario is Strong Coupling Compact Quantum 
Electrodynamics 
(cQED) field theory. This gauge theory describes  the so-called $\pi$-flux RVB state in 
the continuum 
limit\cite{affleck} (see also \cite{Lee},\cite{Rantner}), under the assumption  that the 
system does 
not develop any spin-Peierls ordering.  The gauge theory  has infinite bare coupling and 
therefore 
is difficult to study. All existing analysis 
relies on the results obtained for the SU(N) or 
Sp(2N) generalizations of this theory based on 
the 1/N-expansion, as in 
\cite{ReadSachdev},\cite{affleck},\cite{Lev},\cite{nagaosa}. 

In this paper  we describe  two D-dimensional (D $>$ 1) models which 
exhibit   
neutral spin-1/2 excitations (spinons) and apparently 
represent realizations of the $\pi$-flux RVB state. 
The paper is organized as follows. In Section 2 we 
introduce the models.  They 
are spatially anisotropic and actually represent
collections of weakly coupled chains. 
In all of these models, 
the naively  strongest interactions between the chains, 
namely
those that couple the antiferromagnetic (with momentum 
close to $\pi$) fluctuations on neighbouring chains, are frustrated
and so can be set to zero in the Hamiltonian. In this case, the physics is governed 
by a 
{\sl marginally} relevant interaction  between the long-wavelength magnetic fluctuations.
In Section 3 we discuss  results we managed to obtain for infinite number of chains.  
 These can be summarized as follows:
\begin{itemize}
\item
 we give a proof of the existence  of spin--1/2 excitations
in the limit of infinite number of chains;
\item
 we  demonstrate that in the low-energy limit the effective spin Hamiltonian decouples 
into two commuting parts describing sectors with different parity;
\item
we show that the models are stable against an antiferromagnetic phase transition;
\item
for the models in question we demonstrate existence of a zero temperature phase 
transition (spontaneous transverse dimerization).
\item
we show that the ground state is degenerate; for periodic transverse 
boundary conditions 
the degeneracy is equal to 2$^{2N -1}$ where $2N$ is the number of chains. 
\end{itemize}

To achieve a more detailed understanding of the model, its spectrum and 
correlation functions, we study some limiting cases. Thus   
in Section 
4 we discuss two exactly solvable cases: two 
and four chains with periodic boundary 
conditions in the transverse direction. In Section 5 we consider another 
solvable case: the model of three chains with open boundary conditions. In all 
these cases magnetic excitations carry spin 1/2, in agreement with the general 
result. Systems with more than two chains also display non-magnetic (singlet) 
modes and a topological order. We show that this order is similar to the one which 
exists in the 1D Quantum Ising model. 

Further, in Section 5 we introduce a two-dimensional model with a hierarchy of 
interactions. We use this, apparently artificial but solvable model as a substitute 
for the original (spatially uniform) one with a hope to extract 
more detailed information about the 
structure of the Hilbert space. The hierarchical model is solved by 
a cluster expansion. The results indicate that about half of the Hilbert space is 
occupied by singlet modes. The ground state remains 
massively degenerate with the ground state entropy proportional to the number of chains. 

Having in view  future numerical simulations for our models,  we discuss in Section 
6 finite--size 
effects for systems of a finite number of chains. In Section 7 we use the 
acquired knowledge to conjecture the 
properties of correlation functions in the thermodynamic limit. Our conclusions are 
summarized in Section 8.

\section{The models}

As in the original suggestion by Anderson \cite{anderson}, 
the element most  essential for our 
construction is  frustration. Another ingredient 
of our approach is strong spatial anisotropy: so far we have only been able to   
deal with models where  the exchange interaction in one direction is much greater than 
in the others.  Thus it is proper to describe our models as  
assemblies of weakly coupled chains. Our results  also suggest 
an alternative approach to Strong Coupling  cQED. 
We will not elaborate on this analogy, though, postponing 
this interesting question for future studies.

The first model is a spin-1/2 Heisenberg magnet on an anisotropic Confederate Flag  
(CF) lattice.   
In what follows, we will be working with a
two-dimensional version of this model. The interaction pattern  
depicted on Fig. 1a; one can easily generalize this construction to  three dimensions.
The CF model Hamiltonian is given by
\bea
H_{ACB} = \sum_{j,n}\left\{J_{\parallel}{\bf S}_{j,n} \cdot {\bf S}_{j+1,n} + 
\sum_{\mu = \pm 1} \left[ J_r {\bf S}_{j,n} + J_d
\left( {\bf S}_{j+1,n} + {\bf S}_{j-1,n} \right) \right] \cdot {\bf S}_{j,n + \mu} 
\right\},
\eea
where ${\bf S}_{j,n}$ are spin-1/2 operators, and 
$J_{\parallel}, J_r , J_d > 0$. Since the interchain couplings ($J_r, J_d$) 
are  much smaller than the exchange along the chains ($J_{\parallel}$)
it is legitimate to adopt 
a continuum description of individual chains. In this description,
the local spin densities are represented as sums of the smooth and  staggered parts:
\be
{\bf S}_{j,n} /a_0 \to   {\bf S}_n (x) = {\bf M}_n(x) + (-1)^j{\bf N}_n(x),
~~~x = j a_0,
\label{continS}
\ee
$a_0$ being the lattice spacing in the chain direction.
The low-energy dynamics of the spin--1/2 Heisenberg antiferromagnet 
\bea
H_{1D} = J_{\parallel}\sum_j({\bf S}_j{\bf S}_{j + 1})
\eea
is 
described by the SU$_1$(2) Wess-Zumino-Novikov-Witten model.  
The latter  Hamiltonian can be written in terms of the so-called 
chiral vector {\it current} operators, 
${\bf J}$ and $\bar{\bf J}$, satisfying the level $k =1$ Kac-Moody algebra 
(this approach has been described in a 
vast number of publications; see for a review \cite{affleck1} or \cite{Book}):
\bea
H_{1D} \to  \frac{2\pi v}{3}\int \rd x\left[:({\bf J}\cdot {\bf J}): + 
:({\bf \bar J}\cdot {\bf \bar J}):\right] + \cdots,
\label{WZNW}
\eea
where $v = \pi J_{\parallel} a_0/2$ is the spin velocity
and the dots stand for
a marginally irrelevant perturbation which will be discarded in what follows.  
It is remarkable that the smooth part of magnetization,
\be
{\bf M} = {\bf J} + \bar{\bf J}, \label{M}
\ee 
and the spin current,
\be
{\bf j} = v({\bf J} - \bar{\bf J}), \label{j}
\ee
are locally expressed in terms of the chiral currents.

In the CF model, the exchange 
is frustrated in the direction perpendicular to the chains 
and can be fine tuned  to make its Fourier transform vanish at 
$q_{\parallel} = \pi$.
This is achieved when the rung and plaquette-diagonal
coupling constants satisfy the relation 
$J_r = 2J_d$, in which case  
the direct interchain interaction between staggered magnetizations, 
${\bf N}_n (x) \cdot {\bf N}_{n+1} (x)$, is completely eliminated. 
The absence of this, strongly relevant perturbation
is the most important property of our model. The triangular lattice is 
excluded from the consideration. 
It was demonstrated in 
\cite{twist} that  the coupling of the staggered parts of magnetization 
cannot be removed completely in that case: parity-breaking
(``twist'') terms surviving the continuum limit 
make the analysis very complicated. 
\begin{figure}[ht]
\begin{center}
\epsfxsize=0.65\textwidth
\epsfbox{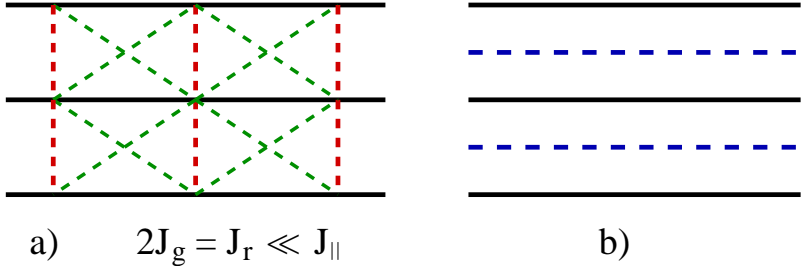}
\end{center}
\caption{a.) Exchange interactions pattern for CF model, b.) Lattice of stripes for 
the Kondo-Heisenberg model.\label{fig:kink}}
\end{figure}

Apart from the collective  spin excitations,
the second  model also involves  charge degrees of freedom. 
It is a quasi-one-dimensional Kondo-Heisenberg model (see Fig.1b) which has 
 been  already discussed  in the context of  theory of  stripes 
\cite{kivelson},\cite{zachar} (see also \cite{white} where the two chain case is 
discussed). It is assumed that neighboring chains have different 
band filling such that metallic chains   
are surrounded by insulating spin-1/2 magnetic chains. In order to eliminate the 
electron tunneling across the magnetic chain, one needs to have magnetic stripes 
consisting of at least three chains. At low energies one can consider such stripe 
as an effective single spin-1/2 chain. Here the frustration 
is dynamical:  the electrons living on metallic chains do not experience 
back-scattering on magnetic excitations due to incommensurability of the Fermi  
wave vectors: $2k_F \neq \pi$. 
It is easy to demonstrate  (see, for instance \cite{kivelson}) that
 the charge excitations in this model 
decouple and remain  one-dimensional 
(this is true at least in the first approximation when one 
discards various virtual processes). On the other hand,
the spin sector is described by the 
same effective Hamiltonian as in the CF model. 

Using the continuum description of individual chains in the
interchain exchange interaction, based on 
the asymptotic representation (\ref{continS}) of the spin operators,
we arrive at the following effective Hamiltonian:
\bea
H = \sum_{n =1}^{2N} \left[H_{1D,n} + \frac{\gamma}{2}\sum_{\mu=\pm 1}({\bf J} + 
\bar{\bf J})_n \cdot ({\bf J} + \bar{\bf J})_{n + \mu}\right].\label{model}
\eea
where $H_{1D}$ is given by Eq.(\ref{WZNW}). Here  $\gamma$ is determined by the 
lattice Hamiltonian; in the CF model 
with interactions chosen as in Fig. 1a we have
$\gamma = 2J_r a_0$.
 For the Kondo-Heisenberg model shown in Fig.1b,  Eq.(\ref{WZNW}) is modified in 
such a way that even (metallic) and odd chains (effective spin-1/2 chains representing 
a stripe of three coupled Heisenberg chains)  
have different velocities. 
This detail, however, does not lead to any qualitative changes.

The model (\ref{model}) is closely related to the gauge theories extensively
studied in the 
context of strongly correlated systems.
This analogy is briefly discussed in  Appendix A. In the rest of this
paper, however, we will follow a different route. Namely, we will
return to Eqs.(\ref{WZNW}), (\ref{model}) and employ the Abelian bosonization
which proves particularly useful in revealing the topological nature of
elementary excitations and determining their quantum numbers.

\section{Proof of existence of the spin-1/2 excitations. Infinite number of chains}

It is well known (see e.g. \cite{Book}) that the chiral SU$_1$(2) currents
can be faithfully represented in terms of holomorphic (left) and antiholomorphic 
(right) scalar fields, $\varphi$ and $\bar\varphi$:
\bea
J^3 _n = \frac{\ri}{\sqrt{2\pi}}\p\varphi_n, ~~&&~~ J^{\pm}_n = 
\frac{1}{2\pi \alpha}\re^{\pm\ri\sqrt{8\pi}\varphi_n}\label{curr-bos1}\\
\bar J^3 _n= -\frac{\ri}{\sqrt{2\pi}}\bar\p\bar\varphi_n, ~~&&~~ {\bar J}^{\pm} _n= 
\frac{1}{2\pi \alpha}\re^{\mp\ri\sqrt{8\pi}\bar\varphi_n}. \label{curr-bos2}
\eea
Here 
$\p,\bar\p = \frac{1}{2} (v^{-1}\p_{\tau} \mp \ri \p_x)$,
$\alpha$ is the ultraviolet cut-off of the bosonic theory, and the
fields $\varphi$ and $\bar\varphi$ are governed by the {\it chiral} Gaussian 
actions \cite{Jackiw},\cite{Moon} (the latter work used this form of bosonization 
in the context of the theory of {\it edge states} in Quantum Hall Effect). 
The bosonized low-energy effective action can then be suitably represented as
\bea
&&S = \sum_{n = 1}^N\int \rd \tau \rd x~ [{\cal L}^+ _n + {\cal L}^- _n 
+ {\cal L}^{int} _n], \label{entire}\\
&&{\cal L}^+ _n = \p_x\varphi_{2n}(\p_x\varphi_{2n} - \ri v^{-1}\p_{\tau}\varphi_{2n}) +
 \p_x\bar\varphi_{2n + 1}(\p_x\bar\varphi_{2n + 1} 
+ \ri v^{-1}\p_{\tau}\bar\varphi_{2n + 1}) +  
\nonumber\\
&& ~~~+ \gamma\sum_{\mu = \pm 1}\left\{(2\pi \alpha)^{-2}\cos[\sqrt{8\pi}(\varphi_{2n} + 
\bar\varphi_{2n + \mu})] + 
\frac{1}{2\pi}\p_x\varphi_{2n}\p_x\bar\varphi_{2n + \mu}\right\}, 
\label{Action}\\
&&{\cal L}^- _n = {\cal L}^+ _n (\varphi \rightarrow \bar\varphi),\label{bosmodel}\\
&&{\cal L}^{int} _n = 
\gamma\sum_{\mu = \pm 1}\left\{(2\pi \alpha)^{-2}\cos[\sqrt{8\pi}
(\varphi_{2n} - \varphi_{2n + \mu})] + 
\frac{1}{2\pi}\p_x\varphi_{2n}\p_x\varphi_{2n + \mu}\right\} 
+ (\varphi \rightarrow \bar\varphi).
\eea
This is the form of the action in which the requirement of vanishing  charge current 
($J_c^{\mu} =0$), imposed in the gauge field RVB approach, is explicitly resolved.

According to the definitions (\ref{curr-bos1}), (\ref{curr-bos2}),
the total spin projection $S^z$ is equal to
\bea
S^z = \frac{1}{\sqrt{2\pi}}\sum_n\int_{-\infty}^{\infty}\rd x \left( \p_x\varphi_n + 
\p_x\bar\varphi_n \right). \label{spinz} 
\eea
This is a general definition which does not assume that the fields
$\varphi$ and $\bar\varphi$ are independent. In fact, the interchain interaction 
that flows to strong coupling in the low-energy limit freezes certain combinations
of the fields with opposite chiralities and thus makes them coupled.

In the first loops the Renormalization Group (RG) equations for the model of infinite number 
of chains coincide with the equations for two chains. At $\gamma > 0$  the 
interaction of currents with different 
chirality in (\ref{Action}), (\ref{bosmodel}) is marginally relevant
and reaches the strong coupling at an energy scale
\be
\Delta = C(v\gamma)^{1/2}\exp(- \pi v/\gamma),\label{scale}
\ee
where $C$ is a number.
At the same time, the (Lorentz-noninvariant)  perturbation  
${\cal L}^{int}$, which is responsible for the velocity renormalization, is irrelevant 
and flows only in the presence of the other interaction. Thus it is likely that
this interaction remains weak even when the relevant one reaches 
the strong coupling. 
For this reason, in the situation when the bare constant $\gamma \ll 1$,
we deem it possible to neglect ${\cal L}^{int}$. 
This results in an approximate Lorentz invariance in  (1+1)-dimensional space-time
which plays an important role in the analysis that follows.
Without  
${\cal L}^{int}$,  the action splits into two independent,
chirally asymmetric sectors,
$S = S^{+} + S^{-}$. 
The two sectors are mapped onto each other 
under the reversal of the chiralities, $\varphi_n \leftrightarrow \bar\varphi_n$,
or  a shift by one lattice spacing in the 
transverse direction. 
Hence the total symmetry becomes  [SU(2)]$_{+}\otimes$[SU(2)]$_{-}$ and  the excitations  
in this model  carry two quantum numbers: spin and parity.

Notice that
the multi-chain effective action (\ref{entire}) features a local Z$_2$ symmetry already 
mentioned for the two-chain problem \cite{allen}: 
as opposed to the original lattice model
which displays global translational invariance, the 
effective Hamiltonian of the continuum theory (\ref{model})
is fully expressed in terms of the vector currents and, therefore,
is invariant under
{\it independent} translations along the chains by 
one lattice spacing. These translations are generated by the shifts
\be
\varphi_n \to \varphi_n \pm \sqrt{\frac{\pi}{2}}, ~~~\bar\varphi_n \to \bar\varphi_n,
\label{transl1}
\ee
or
\be
\varphi_n \to \varphi_n , ~~~\bar\varphi_n \to \bar\varphi_n \pm \sqrt{\frac{\pi}{2}}.
\label{transl2}
\ee
The development of strong-coupling regime in the low-energy limit 
leads to a spontaneous breakdown of this symmetry. Indeed, in the ground state
the field combinations $\varphi_{n} + \bar\varphi_{n + \mu}$
get frozen at the values (see Appendix B for the details)
\be
\varphi_{n} + \bar\varphi_{n + \mu} = \sqrt{\frac{\pi}{8}}
+ \sqrt{\frac{\pi}{2}} m^{(\mu)}_{n},
~~~m^{(\mu)}_{n} = 0, \pm1, \pm2 , ... \label{vac-fields}
\ee
For an even number of chains and with periodic boundary conditions in the 
transverse direction the integers $m^{(\mu)}_{n}$ satisfy
\be
\sum_{n} m^{(+)}_{n} = \sum_{n} m^{(-)}_{n}. \label{integers} 
\ee
The sets of integers $\{ m^{(\mu)}_{n}  \}$ subject to the condition
(\ref{integers}) 
label different
degenerate ground states, each of them breaking the local translational
symmetry (\ref{transl1}), (\ref{transl2}). The existence of stable degenerate
minima of the potential implies the existence of
massive topological excitations, solitons, interpolating between adjacent vacua.
Using (\ref{spinz}) and (\ref{vac-fields}) one can represent the spin
of a given state as 
\be
S^z = \frac{1}{2} \sum_n \left[ m^{(+)}_{n} (\infty)
- m^{(+)}_{n} (- \infty)\right]. \label{spinz1}
\ee
Thus, the minimal nonzero value of the spin is 1/2. This is the spin
carried by the topological solitons.

We emphasise that this statement is based on the exact 
symmetries of  multi-chain 
Hamiltonian (\ref{entire}) and does not depend on  the adopted approximations.  
Within the approximation that neglects ${\cal L}^{int}$ there are solitons and
antisolitons with a well defined (+) and (--) parity.

We would like to briefly comment on the above results.
The factor $\sqrt{8\pi}$ that appears in the arguments 
of the cosines in Eqs.(\ref{Action}),  (\ref{bosmodel}),
representing a marginal current-current interchain interaction, is crucial. 
Recalling the bosonization rules for the staggered magnetization of 
a single spin-1/2 chain (see e.g. \cite{Book})
\be
{\bf N}_n \sim \left(\cos \sqrt{2\pi}(\varphi_n - \bar\varphi_n),~
\sin \sqrt{2\pi}(\varphi_n - \bar\varphi_n),~
- \sin \sqrt{2\pi}(\varphi_n + \bar\varphi_n),
\right) \label{stag-bos}
\ee
we realize that interchain coupling between staggered magnetizations
would give rise a strongly relevant perturbation containing cosines with $\sqrt{2\pi}$.
If such terms were present in the action, the period of the potential 
and, hence,
the topological quantum numbers of the excitations, would be doubled
as compared to our case, and thus the spinons would be confined.
However, by construction these terms do not appear in  our models,
and this is why the interchain interactions do not lead to spinon confinement. 
This argument does not require neglecting $L_{int}$ because the latter term  
also contains only $\sqrt{8\pi}$ cosines. 

Thus we are confident that in the 
models in question  spin-1/2 excitations do exist. We also may rest assured 
that these excitations do not remain confined to individual chains, as it 
happens, for instance in  the so-called 
'sliding' Luttinger liquid phase in arrays of crossed chains
\cite{slide1}, \cite{slide2}. 
In that latter case,  the spinons survive due to irrelevance of  the interchain 
interactions. 
Contrary to that scenario, in our case the interchain coulings 
remain relevant making the system truly multi-dimensional.

\subsection{A failure of the 'most obvious'  solution}

The following calculation suggests that freezing of the fields does not imply that 
one can expand around the vacuum configurations. Naively, one would adopt
a self-consistent harmonic approximation (SCHA)
in which the cosine term 
is expanded around the minimum:
\bea
\cos[\sqrt{8\pi}(\varphi_{n} + \bar\varphi_{n + \mu} )] \approx \mbox{const} + \frac{1}{2}
\rho[\delta(\varphi_{n} + \bar\varphi_{n + \mu})]^2 .
\eea
Here $\delta$ denotes a deviation from the vacuum value. SCHA assumes
that  the transverse stiffness $\rho$ should be  determined self-consistently:
\bea 
\rho = \frac{2\gamma}{\pi\alpha^2}\la
\exp[\ri\sqrt{8\pi}(\varphi_{n} + \bar\varphi_{n + \mu} )]\ra . \label{sf}
\eea
We shall demonstrate, however, that such procedure yields zero $\rho$. 

 The effective action for fluctuations around the minima  is 
(here we set $v=1$)
\bea
S = \sum_p(\alpha^* _p, \beta^* _p)\left(
\begin{array}{cc}
q(q + \ri\omega) + \rho, & \rho\cos k\\
\rho\cos k, & q(q - \ri\omega) + \rho
\end{array}
\right)
\left(
\begin{array}{c}
\alpha_p\\
\beta_p
\end{array}
\right) ,
\eea
where 
$\alpha = \delta\varphi, ~\beta = \delta\bar\varphi$,
$p = (\omega,q,k)$, $q$ and $k$ being the longitudinal and transverse momenta, respectively. 
Self-consistency condition (\ref{sf}) becomes
\bea
\ln\rho \sim 
&&- [G_{\alpha\alpha} + G_{\beta\beta} + 2G_{\alpha\beta}](\tau=0,x =0) \sim
\int\rd\omega\rd q\rd k\frac{q^2 + \rho(1 - \cos k)}
{\omega^2q^2 + (q^2 + \rho)^2 - \rho^2\cos^2k} \nonumber\\
&&\sim -\int \frac{\rd k\rd q}{q}\frac{q^2 + \rho(1 - \cos k)}{\sqrt{(q^2 + \rho)^2 - 
\rho^2\cos^2k}}
\eea
Since the latter integral diverges, the suggested naive  
solution is not self-consistent. Its  failure originates from  the chiral 
structure of the Gaussian part of action (\ref{Action}) which includes  
the first power of the time gradient. For the conventional (non-chiral) 
Gaussian model a similar solution would be self-consistent. One immediate 
consequence of it  would be the appearance of an average staggered magnetization. 
Thus the present calculation demonstrates {\it stability against antifferomagnetism}.

\subsection{Order parameters}

As we have already seen,
the fact that the combinations of the fields $\varphi_{2n} + \bar\varphi_{2n+\mu}$
and  $\bar\varphi_{2n} + \varphi_{2n+\mu}$ get locked at
$\sqrt{\frac{\pi}{8}} + \sqrt{\frac{\pi}{2}} n$, $(n \in Z)$ implies that
the ground state of the system is characterized by a spontaneously broken 
local translational invariance
(\ref{transl1}), (\ref{transl2}).
The {\it most obvious} order parameters (OP) in the two sectors of the model
(\ref{entire}), whose structure 
directly follows from the form of the interaction,
are given by 
\bea
&&{\cal T}_{\mu}^{(+)} = 
\ri\la\exp[\ri\sqrt{2\pi}(\varphi_{2n} + \bar\varphi_{2n + \mu})]\ra, \nonumber\\
&&{\cal T}_{\mu}^{(-)} = 
\ri\la\exp[\ri\sqrt{2\pi}(\bar\varphi_{2n} + \varphi_{2n + \mu})]\ra \label{OP}
\eea
The above  OPs are real
(see Eq.(\ref{vac-fields})),
defined on the links of the lattice and  are 
non-local in terms of original spins. Moreover, since the OPs  include chiral fields, they 
are expressed (nonlocally!) in terms of magnetization density and {\it spin currents}. 
As is shown in Appendix B, for different pairs of neighboring chains, 
the signs of the order parameters 
are uncorrelated.
This leads to a massive gound state degeneracy 2$^{2N -1}$ 
(recall that $2N$ is the total number of chains).

The non-locality of the order parameter suggests a possibility  of  {\it topological} 
order.  This supports the original idea by Wiegmann that topological order exists 
in RVB states  \cite{toporder},\cite{Wiegmann}.   
 Characteristic features of such order is non-locality of the order parameter, 
ground state degeneracy (in a closed system) and the existence of edge states 
(in an  open system). The ground state degeneracy depends on the topology of the manifold 
(hence the name 'topological'). A relative stability of the state with respect to 
perturbations is protected by the spectral gap. One well known example of topologically 
ordered states is provided by incompressible Fractional Quantum Hall states 
(see the discussion in \cite{wen} and references therein). In our case the degeneracy 
of the ground state is exponential in the number of chains. This fact alone indicates 
that  these  models have a topological order  of a different type. This order is closely 
related to the one existing in 2D Ising model, 
multi-channel Kondo models  and models with scattering matrices of the RSOS-type. 

The first examples of non-local  order parameters are given by the disorder field  
of 2D Ising model \cite{kadanoff1} and the dual field exponentials in 2D 
XY model\cite{jose},\cite{pasha}. These fields acquire non-zero expectation values in 
disordered 
phases of the corresponding models. 
According to   \cite{denNijs},\cite{Girvin},\cite{Kohmoto}, \cite{SNT},\cite{Book}, 
such order parameters (also called 'string' order parameters) exist in various massive 
phases of 1D spin liquids. All these liquids are related to either spin S=1 Heisenberg chain 
or two-leg spin 1/2 Heisenberg ladders. As  was demonstrated in 
\cite{AT},\cite{SNT},\cite{NT} (see also \cite{Book}), after the Wick rotation 
these systems can be described as a collection  of 2D Ising models. The evidence that 
the order is really topological was 
presented by Affleck \cite{af}, who  proved that 1D spin liquids have edge, or rather end,  
states: open chains have free spin-1/2 states located at the chain's ends.

 We postpone a more detailed discussion of the issue of topological nature of the order 
till the further Sections. 

Order parameters (\ref{OP}) change their sign on soliton configurations and, therefore, 
vanish 
at finite temperatures when the densities of  solitons and antisolitons  are  finite and 
{\it equal}. This mechanism is  a sufficient (though not a necessary) one  to destroy 
the order 
parameter at finite $T$. The sign changes may also originate from fluctuations in 
non-magnetic channels. Later we will discuss  an exactly solvable example of four coupled 
chains in which the  latter  scenario is realized. 
The temperature dependence of the correlation 
length depends on what mechanism is realized. If the order parameter is destroyed only by 
(anti)solitons, the correlation length is exponentially large in $1/T$. This is because 
the  topological excitations have
a finite  spectral  gap and therefore  their density  is 
exponentially small: $n \sim \exp(- \Delta/T)$. 
In this case, the correlation length is defined as the average spacing
between the solitons:
$\xi \sim n^{-1/D}$ ($D$ is the space dimensionality).

 As we shall see, the singlet excitations either have a very small gap or 
even no gap at 
all (we do not know whether the latter is true). Then the order parameter is governed by  
another correlation length quite distinct from the above. The density of   non-magnetic 
excitations  above their gap  depends on $T$ as a power law. Such  situation would 
correspond to a quantum critical point with a power law correlation length. 

Turning to physical fields, we note that the products of the staggered magnetizations on 
neighboring chains representing the transverse dimerization order parameters (see
Appendix B) are  expressed  as  bilinear 
combinations of the true  order parameters:   
\bea
&&\la N_{2n}^zN_{2n + \mu}^z \ra =  \frac{1}{8\pi a_0}\la\cos\left\{\sqrt{2\pi}
(\varphi_{2n} + \bar\varphi_{2n + \mu} +\bar\varphi_{2n} + \varphi_{2n + \mu})\right\}\ra 
=  \frac{1}{4}[{\cal T}_{\mu}^{(+)}{\cal T}_{\mu}^{(-)} + c.c],\\
&&\la N_{2n}^xN_{2n + \mu}^x\ra  = \la N_{2n}^yN_{2n + \mu}^y\ra \nonumber\\
&&= \frac{1}{8\pi a_0}\la\exp[\ri\sqrt{2\pi}(\varphi_{2n} + \bar\varphi_{2n + \mu})]
\exp[-\ri\sqrt{2\pi}(\bar\varphi_{2n} + \varphi_{2n + \mu})]\ra  \nn\\
&&= \frac{1}{4}\{{\cal T}_{\mu}^{(+)}[{\cal T}_{\mu}^{(-)}]^* + c.c.\}.
\eea
Since ${\cal T}^* = {\cal T}$ we have  
\[
\la N^a_n N^b_{n + \mu}\ra \sim \delta^{ab},
\]
implying  that the SU(2) symmetry remains unbroken in the ground state.

Now let us consider the magnetic correlation functions.  The order parameter field of 1D 
spin-1/2 chain is not just a staggered magnetization; it is a 2$\times$2 SU(2) matrix 
$\hat g$ whose entries include not only vector components of the magnetization $n^a$ 
($a = x,y,z$), but also the dimerization field  
$\epsilon = (-1)^n({\bf S}_n{\bf S}_{n + 1})$:
\bea
\hat g(x) = \epsilon(x)I + \ri\s^a N^a(x).
\eea
It can also be represented in a factorized bosonic form:
\bea
\hat g_{\s\s'} = \frac{1}{\sqrt 2}C_{\s\s'}:\exp[-\ri\sqrt{2\pi}
(\s\varphi + \s'\bar\varphi)]: \equiv C_{\s\s'}z_{\s}\bar z_{\s'},\nonumber\\
C_{\s\s'} = \re^{\ri\pi(1 - \s\s')/4},
\eea
where 
\bea
 z_{\s} = \exp[\ri\s\sqrt{2\pi}\varphi], ~~ \bar z_{\s} = \exp[-\ri\s\sqrt{2\pi}
\bar\varphi],  ~~(\s = \pm 1).
\eea
As follows from the factorization of the low-energy effective action, if we consider 
correlation functions on chains with the same  parity (for instance,  both even)   
$z$ correlate only with $z$'s and $\bar z$ with $\bar z$. For Green's functions 
on chains with different parity $z$ correlate with $\bar z$ and not with $z$. Therefore  
the  correlation functions of the staggered magnetizations and energy densities factorize: 
\bea
&&\la g_{ab}(\tau,x;2n)g^+_{cd}(0,0;2m)\ra = \la z_a(\tau,x;2n) z^*_d(0,0;2m)\ra\la 
\bar z_b(\tau,x;2n) \bar z^*_c(0,0;2m)\ra \nonumber\\
&&\equiv \delta_{ad}\delta_{bc}
G_{nm}(\tau,x)G_{nm}(\tau, -x),\label{factor1}\\
&&\la g_{ab}(\tau,x;2n)g^+_{cd}(0,0;2m +1)\ra = \la z_a(\tau,x;2n) 
\bar z^*_c(0,0;2m)\ra\la \bar z_b(\tau,x;2n) z^*_d(0,0;2m +1)\ra  \nonumber\\
&&\equiv \delta_{ac}\delta_{bd}D_{nm}(\tau,x)D_{nm}(-\tau,x). \label{factor2}
\eea

Various components of the $z$-s 
are  conveniently  assembled into  vectors and co-vectors:
\bea
\left(z^*_{\s,2n}, \bar z^*_{\s,2n +1}\right), ~~\left(
\begin{array}{c}
z_{\s,2n}\\
\bar z_{\s,2n +1}
\end{array}
\right)
\eea
for the (+) sector and 
\bea
\left(z^*_{\s,2n+1}, \bar z^*_{\s,2n}\right), ~~ 
\left(
\begin{array}{c}
z_{\s,2n+1}\\
\bar z_{\s,2n}
\end{array}
\right)  
\eea
for the (--) sector. 
Then the corresponding correlation functions become matrix elements of the 
2$\times$2
Green's function:
\bea
\hat G = \left(
\begin{array}{cc}
G(\omega,k^z; k_{\perp}) & D(\omega,k^z; k_{\perp})\\
D(- \omega, k^z; k_{\perp} ) & G(\omega, - k^z,k_{\perp})
\end{array}
\right)\label{stag2}
\eea
An important fact is that both the function $D$ and $k_{\perp}$-dependence of $G$ are 
 non-perturbative effects: they do not appear  in any  order of perturbation theory in 
$\gamma$ and, as we shall see later, are proportional to $\exp(- \pi v/\gamma)$.

\section{Exactly solvable cases: two and four chains.}

 From now on we will  discuss  the (+) and (--) 
sectors  independently. To distinguish 
between the amplitude of the RVB order parameter and various mass gaps, we shall denote 
spectral gaps by the letter $m$. 

To check  the ideas outlined in the previous Section, we shall discuss 
some exactly solvable cases.  
Consider a two-dimensional array of an  even number (2N) of chains  with  
periodic boundary conditions. In this case we can  arrange chains in pairs. 
Exact solution is possible in two cases: N = 1 and N =2.

 Throughout this Section we shall use the so-called form-factor approach to calculate 
leading asymptotics of the 
correlation functions. This method has been widely applied to massive integrable field 
theories and there are excellent review articles \cite{Babujian} and even 
books \cite{Smirnovbook} about it. The form-factor approach uses information extracted 
from the scattering theory and symmetry properties of 
 physical operators to calculate matrix 
elements of these operators between various exact eigenstates. To estimate the correlation 
functions one uses the Lehmann expansion. This approach is well defined for theories with   
massive excitations. In one dimension, for reasons of kinematics, matrix elements of 
multi-particles states become smaller and smaller when the number of excited particles 
increases. This feature greatly improves  the convergence  of  the Lehmann expansion and 
allows one to obtain an accurate description of the correlation functions with only a few 
(sometimes even one!) matrix elements involved.

\subsection{N =1 (two chains)}
 
 Though the number of chains is really small, this case contains certain interesting 
features 
which may shed  light on more general cases. 

 For N =1 each  sector with a given parity  is equivalent to the isospin sector of the  
SU(2) invariant Thirring model, as it was 
shown in  \cite{allen}. The excitations 
are massive solitons with spin 1/2. There are no other modes (singlet ones), as it must be 
for the case of large N. For this reason this case is not representative, as we have 
already said. The  massive modes correspond to amplitude fluctuations of $\Delta$. 

 The correlation functions of chiral bosonic exponents for the Thirring model can be 
extracted 
from the recent paper by Lukyanov and Zamolodchikov \cite{LukZam} (see also \cite{singleP}). 
In particular, for the correlation functions of spinon fields (\ref{stag2}) and for the 
correlators of staggered magnetizations we have 
\bea
&&G(\tau,x) = \la\re^{\ri\sqrt{2\pi}\varphi(\tau,x)}\re^{-\ri\sqrt{2\pi}\varphi(0,0)}\ra 
\sim \frac{1}{\sqrt{\ri v\tau - x}}\exp(- m\rho_{12})\nonumber\\
&&\la\re^{\ri\sqrt{2\pi}\Phi_1(\tau,x)}\re^{-\ri\sqrt{2\pi}
\Phi_1(0,0)}\ra \sim \frac{1}{\rho_{12}}\exp(-2m\rho_{12})\nonumber\\  
&&D(\tau,x) = \la \re^{\ri\sqrt{2\pi}\varphi(\tau,x)}\re^{\ri\sqrt{2\pi}
\bar\varphi(0,0)}\ra \sim m^{1/2}K_0(m\rho_{12})\nonumber\\
&&\la\re^{\ri\sqrt{2\pi}\Phi_1(\tau,x)}\re^{\ri\sqrt{2\pi}\Phi_2(0,0)}\ra = 
Z_0 m K_0^2(m\rho_{12}) \label{interchain} 
\eea
where $\rho = \sqrt{\tau^2 + (x/v)^2}$ and the mass gap $m = \Delta$ given by 
Eq.(\ref{scale}). 
In writing down these  formulae we used the single mode approximation for the 
spinon operators (for the details  we again refer the reader to  \cite{LukZam}):
\bea
&&z^+_{\sigma}(\tau,x) = Z_0^{1/2}m^{1/4}\int \rd\theta \re^{\theta/4}
\re^{- \ri x m\sinh\theta}[
\re^{-m\tau\cosh\theta}\hat Z^+_{\s}(\theta) + \re^{m\tau\cosh\theta}\hat 
Z_{-\s}(\theta)]\\
&&\bar z_{\sigma}(\tau,x) = Z_0^{1/2}m^{1/4}\int \rd\theta \re^{-\theta/4}
\re^{ - \ri x m\sinh\theta}[
\re^{-m\tau\cosh\theta}\hat Z^+_{\s}(\theta) + \re^{m\tau\cosh\theta}\hat 
Z_{-\s}(\theta)]\label{single}
\eea
where  $Z_{\pm}$($Z^+_{\pm}$) are annihilation (creation) operators of solitons and 
antisolitons and $Z_0$ is a numerical constant. 
Approximation (\ref{single}) fails at small distances; inclusion of terms with multiple 
soliton production removes the logarithmic singularity of $K_0$.

 The two-point correlation function of currents were calculated in \cite{controzzi}.

 Eq.(\ref{interchain}) supports the earlier claim made  at the end of the previous section, 
that the interchain correlation function of the staggered magnetization is non-analytic 
in $\gamma$ and thus  vanishes in all orders of perturbation theory.  

\subsection{N =2 (four chains)}

In the N =2 case the interaction can be written as 
\bea
({\bf J}_1 + {\bf J}_3)(\bar{\bf J}_2 + \bar{\bf J}_4) + ({\bf J}_2 + 
{\bf J}_4)(\bar{\bf J}_1 + \bar{\bf J}_3) 
\eea
A  sum of two $k =1$ SU(2) currents is the $k=2$ current; moreover, according 
to \cite{FatZam} the sum of two SU$_1$(2) WZNW models (the central charge 2) 
can be represented as the SU$_2$(2) WZNW model with central charge 3/2 and plus one
 massless Majorana fermion (a critical Ising model) with central charge 1/2.   
Using  the results of \cite{FatZam} we   rewrite the entire Hamiltonian density 
(\ref{WZNW}) as follows (here only the  (+)-parity part is written):
\bea
&&{\cal H}_+ = {\cal H}_{massless} + {\cal H}_{massive}\nonumber\\
&&{\cal H}_{massless} =  -\frac{\ri v}{2}\chi_{0}\p_x\chi_{0} + 
\frac{\ri v}{2}\bar\chi_{0}\p_x\bar\chi_{0}\label{Ising}\\
&&{\cal H}_{massive} = \frac{\pi v}{2}(:{\bf I}\cdot {\bf I}: + 
:\bar{\bf I}\cdot \bar{\bf I}:) + \gamma{\bf I}\cdot \bar{\bf I} = 
\frac{\ri v}{2}(- \chi^a\p_x\chi^a + \bar\chi^a\p_x\bar\chi^a) - 
\frac{\gamma}{2}(\chi^a\bar\chi^a)^2 \label{O3} 
\eea
where $a = 1,2,3$ and 
\[
{\bf I} = {\bf J}_1 + {\bf J}_3, ~~ \bar{\bf I} = \bar{\bf J}_2 + \bar{\bf J}_4
\]
\bea
{\bf J}_{1,3} = \frac{\ri}{2}\left\{\pm \chi_0\vec\chi + \frac{1}{2}
[\vec\chi\times\vec\chi]\right\}, ~~ 
\bar{\bf J}_{2,4} = \frac{\ri}{2}\left\{\pm \bar\chi_0\vec{\bar\chi} + \frac{1}{2}
[\vec{\bar\chi}\times\vec{\bar\chi}]\right\}\label{curr2}
\eea
 The fields $\chi, \bar\chi$ stand for real (Majorana) fermions.  

 Eq.(\ref{Ising}) describes a critical Ising model; the corresponding  excitations 
 are  gapless and non-magnetic; they  appear in the  sectors with both parities. 
The criticality is an artefact of the periodic boundary conditions in the 
transverse direction. 

 The massive sector can be described using  three equivalent representations. 
One of those is the SU$_2$(2) WZNW model perturbed by a current-current interaction 
(the ultraviolet central charge $C_{UV} = 3/2$). Here the fundamental field is the SU(2) 
matrix $\hat g$. Another representation is  the O(3) Gross-Neveu (GN) model where 
fundamental fields are Majorana fermions (\ref{O3}). The third representation is three 
critical Ising models coupled by the energy density operators $\epsilon_i$:
\bea
S = \sum_{i=1}^3S_{Ising}(\s_i) - \gamma \sum_{i > j}\int\rd\tau\rd x~ \epsilon_i\epsilon_j
\eea
In the latter case the fundamental fields are order parameter fields of the Ising models or, 
via  the duality relationships, the corresponding disorder parameter fields. 

An external magnetic field  couples to the sum of all currents and therefore only to the 
fermions of the O(3) GN model:
\bea
{\cal H}_{magn} = \ri{\bf h} \cdot ([\vec\chi\times\vec\chi] 
+ [\vec{\bar\chi}\times\vec{\bar\chi}])
\eea
Since this model has a spectral gap, the net magnetization does not appear until the 
magnitude 
of the field reaches the value of the gap.

The Thermodynamic Bethe Anzats equations for O(3) GN model were obtained by one of the 
authors \cite{SUSY}; some of the correlation functions were calculated by 
Smirnov \cite{Smirnov}. 
A very illuminating discussion of the symmetries and the excitation spectrum of this 
model 
can be found in the recent paper by Fendley and Saleur \cite{Fendley}.

The  massive magnetic excitations are particles with a non-Abelian statistics, as in the 
Pfaffian state of Quantum Hall effect \cite{Fradkin}. 
An unusual  fact about the O(3) GN model is that its  ground state is  doubly degenerate. 
This degeneracy is of {\it topological} nature being related to properties of 
{\it zero modes} formed on solitons.   
The picture simplifies drastically  if one takes into account that 
O(3) GN model is equivalent to  Supersymmetric sine-Gordon model (SUSY SG) \cite{SUSY}
\bea
{\cal L} = \frac{1}{2}(\p_{\mu}\Phi)^2 + \frac{\ri}{2}\bar\chi\gamma_{\mu}\p_{\mu}\chi 
+ M^2\sin^2(\beta\Phi) + \ri M\bar\chi\chi\cos(\beta\Phi)
\eea
 with $\beta^2 = 4\pi$. For SUSY SG model  there is a semiclassical limit   $\beta^2 \ll 1$, 
where things are easier to grasp. In this limit the spectrum is determined by the 
minima of  the bosonic part of the potential. There are  two sets of minima: ones where 
$\cos(\beta\Phi) = 1$ and ones where it is --1. This sign difference does not affect 
 the energy spectrum of the Majorana fermion but affects 
expectation values of the 
Ising model order and disorder parameters. Therefore the vacua with different sign of 
$\cos(\beta\Phi)$ are physically distinguishable. It is clear that the fact of this 
degeneracy is related to (i) odd number of Majorana fermions species present and (ii) 
the existence of solitons. Both these features are very general and are likely to survive 
in a  finite parameter domain. The described  situation should  
be opposed to what happens in  the conventional sine-Gordon model, where the  potential is 
just $\cos(\beta\Phi)$ and the degeneracy between its minima  is lifted in the ground state. 

 There are two descriptions of the  spectrum. The first one was suggested by Zamolodchikov 
\cite{Zamolodchikov}. 
This description abandons the notion of asymptotic particle states and, therefore, 
leads to difficulties when being applied to
the correlation functions. The second approach 
 was suggested in \cite{bazhanov} and used in \cite{Smirnov} to calculate the correlation 
functions.  In this approach  solitons are treated as particles  carrying   two quantum 
numbers -- an SU(2) spin $\s =\uparrow,\downarrow$ and an isotopic number $p = \pm $.  
The isotopic part of the  K-soliton Hilbert space is  truncated, that is not all 
combinatorically possible 2$^K$ states are allowed. 
The number of allowed states is $2^{[K/2]}$; in fact, the isotopic space of $K$ solitons is 
isomorphic to the irreducible representation of $K$-dimensional Clifford algebra 
(in other words, one may say that each soliton carries a $\gamma$ matrix - 
a zero mode of the Majorana fermion). In particular, the two-soliton wave function must 
be an isotopic singlet.

 This approach allows one to use the conventional scattering matrix description at the 
end projecting out unwanted states. The two-particle S-matrix is given by the tensor product
\bea
S(\theta_{12}) = S_{ITM}(\theta_{12})\otimes S_{4\pi}(\theta_{12})\label{Sm} \label{smatr}
\eea
where the first S-matrix corresponds to the SU(2) invariant (isotropic) Thirring model 
and second one is the S-matrix of solitons of the sine-Gordon model with a particular 
anisotropy $\xi = 4\pi$ (in Smirnov's notations 
$
\xi = \pi\beta^2/(8\pi - \beta^2)
$).

 The order parameters are  local operators both in the Ising and the WZNW representations:
\bea
&&{\cal T}_{12} = \exp[\ri\sqrt{2\pi}(\varphi_1 + \bar\varphi_2)] = 
\re^{\ri\sqrt{\pi}\Phi_+}\re^{\ri\sqrt{\pi}\Phi_-}\nn\\
&&~~~ = (\s_0\s_3 + \ri\mu_0\mu_3)(\s_1\s_2 
+ \ri\mu_1\mu_2) = \s_0{\mbox{Tr}}\hat g  + \ri\mu_0{\mbox{Tr}} \hat g^+\nonumber\\
&&{\cal T}_{14} = \exp[\ri\sqrt{2\pi}(\varphi_1 + \bar\varphi_4)] = 
\re^{\ri\sqrt{\pi}\Phi_+}\re^{\ri\sqrt{\pi}\Theta_-} = (\s_0\mu_3 + 
\ri\mu_0\s_3)(\s_1\s_2 + \ri\mu_1\mu_2)  
\eea
where $\s_i,\mu_i$ are order and disorder parameters of the corresponding Ising models. 
The spontaneous mass generation freezes $\s_i, ~~(i = 1,2,3)$, but  leaves $\s_0,\mu_0$ 
critical. Therefore the  most singular parts of the above order parameters 
\bea
{\cal T}_{12} \sim \s_0, ~~ {\cal T}_{14} \sim \mu_0
\eea
still have power law correlations  at $T =0$. The correlation length at $T \neq 0$ is 
$\xi \sim 1/T$ and is not determined by the solitons -- a possibility we have mentioned 
in Section IIIB. 

 The chiral exponents  entering in the expressions for the stagerred magnetization are 
non-local:
\bea
\exp[\ri\sqrt{2\pi}\varphi_1] = C_{ab}\sigma^a(z)g^b(z,\bar z) 
\eea
where $\s^a$ are chiral {\it vertex operators} of the critical Ising model 
(see \cite{seiberg} for details). The operators 
$g$ in the ultraviolet limit become vertex operators  of the 
S=1/2 tensor operators  of $k=2$ WZNW model.

 In order to understand the subsequent calculations of the correlation functions the 
reader have to keep in mind two principal facts about excitations in integrable models. 
The fact number one is that these excitations cannot be unambiguously classified as 
fermions, bosons, semions {\it etc.}. Their commutation relations are determined by 
their S-matrix and therefore depend on the momenta of the particles (see, for 
example\cite{ZamZam}). The fact number two is that the creation and annihilation 
operators of elementary excitations in integrable theories are usually strongly 
non-local in terms of the bare creation and annihilation operators. Hence these 
excitations are  extended objects and therefore 
do not belong to any particular representation of the Lorentz group. Therefore their 
Lorentz spin is not fixed and  the same operator  can represent physical fields with  
different Lorentz spin depending on circumstances.   
\medskip

{\bf Correlation functions of the currents}
\medskip

According to Eq.(\ref{curr2}) there are two parts in a given  current operator,
$J_-$ and  
$I$. The first one containing  a product of the gapless fermion $\chi_0$ and  the O(3) 
GN model fermion is not conserved. A GN fermion  is a convolution of two solitons, each 
of them entering  with Lorentz spin 1/4 to make the total spin 1/2. Superficially it may 
appear that the second term (the conserved current) is a convolution of four solitons, 
but this is incorrect: the minimal matrix element contains two solitons which enter here 
with  Lorentz spin 1/2. Since $\chi_0$ remains massless, the leading asymptotics of the 
correlation function of non-conserved currents $\la J^-J^-\ra$ with the threshold at 
$s = 2m$ ($s^2 = \omega^2 - (vq)^2$) 
exists only for the currents with the same chirality (here we put $v =1$):
\bea
&&\la J^a_-(\tau,x)J^b_-(0,0)\ra = \delta_{ab}(\tau + \ri x)^{-1}\times\\
&&\int\rd\theta_1\rd\theta_2|F_1
(\theta_{12})|^2\re^{(\theta_1 + \theta_2)/2}\exp[- m\tau(\cosh\theta_1 + \cosh\theta_2) 
+ \ri m x(\sinh\theta_1 + \sinh\theta_2)]\nonumber 
\eea
which gives
\bea
\frac{(\tau - \ri x)}{(\tau + \ri x)\sqrt{\tau^2 + x^2}}\int\rd\theta 
|F_1(\theta)|^2K_1(2m\rho\cosh\theta)
\eea
where $F_1$ can be calculated. The Fourier transform 
\bea
&&\Im m\la J^{-}J^{-}\ra_{(\omega,q)}= \left(\frac{\omega + q}{\omega - q}\right)^2 
L(s^2), ~~\Im m\la \bar J^{-}\bar J^{-}\ra_{(\omega,q)}= 
\left(\frac{\omega - q}{\omega + q}\right)^2 L(s^2), \nonumber\\
&&L(s^2) = s^2\int\frac{\rd\theta}{\cosh^3\theta}\theta_H(s^2 - 4m^2\cosh^2\theta)
|F_1(\theta)|^2
\eea
where $\theta_H(x)$ is the Heaviside function.

 Now let us consider the pair correlation function of the conserved  O(3) currents 
$I = J_1 + J_3, ~~\bar I = \bar J_2 + \bar J_4$. Their matrix elements were calculated 
in \cite{Smirnov}. In the scheme adopted in that paper a soliton carries  two quantum 
numbers - an SU(2) spin $\s =\uparrow,\downarrow$ and an isotopic number $p = \pm $. 
The isotopic part of the  multisoliton Hilbert space is truncated such that  the two 
soliton wave function must be an isotopic singlet.  Due to the  SU(2) symmetry, it is 
sufficient to have an expression for one component of the current, for example $I^3$. 
The matrix element into a state of two solitons with rapidities $\theta_1,\theta_2$ 
and quantum numbers $(\s,p)_1,(\s,p)_2$ is given by  
\bea
&&\la\theta_1,(\s_1,p_1);\theta_2,(\s_2,p_2)|I^z|0\ra = d\re^{(\theta_1 + \theta_2)/2}
\frac{\coth(\theta_{12}/2)\zeta_{\infty}(\theta_{12})
\zeta_{4}(\theta_{12})}{(\theta_{12} + \ri\pi)\cosh[\frac{1}{8}(\theta_{12} + \ri\pi)]}
\times\nonumber\\
&&[|\uparrow\downarrow\ra + |\downarrow\uparrow\ra]\otimes[|+-\ra - 
|-+\ra]
\eea
where 
\bea
&&\zeta_{k}(\theta) = \sinh(\theta/2)\tilde\zeta_k(\theta), \nonumber\\
&&\tilde\zeta_k(\theta) = \exp\left[\int_0^{\infty}\rd x
\frac{\sin^2[x(\ri\pi + \theta)]\sinh[\pi(1 - k)x/2]}{x\sinh\pi x\cosh(\pi x/2)
\sinh(\pi kx/2)}\right]
\eea
and 
\[
d = \frac{1}{64\pi^3\zeta_{\infty}(\ri\pi)\zeta_4(\ri\pi)}
\] 
Thus 
the leading asymptotics of  $\la I^a I^b\ra$ and $\la I^a \bar I^b\ra$ are 
\bea
&&\la I^a(\tau,x)I^b(0,0)\ra = 2d^2\delta_{ab}\times\\
&&\int\rd\theta_1\rd\theta_2|F_2(\theta_{12})|^2
\re^{(\theta_1 + \theta_2)}\exp[- m\tau(\cosh\theta_1 + \cosh\theta_2) + 
\ri mx(\sinh\theta_1 + \sinh\theta_2)]\nonumber\\
&&\la I^a(\tau,x)\bar I^b(0,0)\ra = \nonumber\\
&&\delta_{ab}2d^2\int\rd\theta_1\rd\theta_2|F_2(\theta_{12})|^2
\exp[- m\tau(\cosh\theta_1 + \cosh\theta_2) + \ri mx(\sinh\theta_1 + \sinh\theta_2)]
\eea
where 
\bea
|F_2(\theta)|^2 = \frac{\sinh^2\theta}{[\cosh(\theta/4) + \cos(\pi/4)]
(\theta^2 + \pi^2)}|\tilde\zeta_{\infty}(\theta)\tilde\zeta_4(\theta)|^2
\eea
For the Fourier transform we have 
\bea
&&\la I^a_{\mu}I^b_{\nu}\ra = \delta_{ab}
(\delta_{\mu\nu} - q_{\mu}q_{\nu}/q^2)D(s^2)\nonumber\\
&&\Im m D(s^2)=\frac{4d^2}{\sqrt{s^2 - 4m^2}}|F_2(\theta)|^2, ~~
\cosh^2(\theta/2) = s^2/4m^2
\eea
such that at the threshold we have the same behavior as for $N =1$:
\bea
\Im m D(s^2) \sim \sqrt{s^2 - 4m^2}
\eea
\medskip

{\bf Single-electron Green's function for the model of stripes}
\medskip

All correlation functions discussed so far exist for both models depicted on Fig. 1. 
Now we are going to discuss the single-electron function which exists only for the model 
of stripes (Fig. 1b). For $N=1,2$ we can calculate the spinon 
Green's function $G$ and find the following expression for the spectral function:
\bea
A_R(\omega,q) \sim \frac{1}{\omega - vq}
\left(\frac{m^2}{\omega^2 - (vq)^2 - m^2}\right)^{3/8}
\eea
and for the tunneling density of states:
\bea
\rho(\omega) \sim \int_0^{\cosh^{-1}(\omega/m)}
\frac{\rd\theta\cosh(3\theta/8)}{(\omega/m - \cosh\theta)^{3/8}}
\eea
\medskip

{\bf Correlation functions of the staggered magnetizations}
\medskip

 As we have stated above, the correlation functions of staggered magnetizations factorize 
into a product of two kinds of correlation functions: $G$ and $D$ 
(see Eqs.(\ref{factor2},\ref{stag2})). The function $D$ is essentially non-perturbative and 
therefore especially interesting. We have demonstrated that for $N =1$ this function is 
non-zero. For $N =2$ with the periodic boundary conditions   this  function vanishes.  
This is related to the fact that for the periodic boundary conditions the singlet sector 
is critical. The operator 
$\exp[\ri\sqrt{2\pi}\varphi]$ contains  vertex operators  $\s_0$; at criticality  
correlation functions of such operators  with vertex operators  of different chirality 
always vanish: $\la\s_0\bar\s_0\ra =0$.

 To understand this better it is convenient to write the staggered magnetizations in 
the appropriate basis. Thus we have 
\bea
&&{\bf N}_{1,3} = \sigma_0^{(1)}\mbox{Tr}[\vec\s(g + g^+)] 
\pm \ri\mu_0^{(1)}\mbox{Tr}[\vec\s(g - g^+)]\nonumber\\
&&{\bf N}_{2,4} = \sigma_0^{(2)}\mbox{Tr}[\vec\s(G + G^+)] 
\pm \ri\mu_0^{(2)}\mbox{Tr}[\vec\s(G - G^+)]\label{staggered}
\eea
where $g, G$ are the SU(2) matrices. The fields $\s_0,\mu_0$ remain critical. 
The fields $g,G$ have to be decomposed into their vertex operators and the interaction 
involves sectors with different chirality. 
 
From Eqs.(\ref{staggered}) we have
\bea
&&\la\la {\bf N}_1{\bf N}_1\ra\ra = \left(\la\la\s_0\s_0\ra\ra + 
\la\la\mu_0\mu_0\ra\ra\right)\la\la\mbox{Tr}[\vec\s g]\mbox{Tr}
[\vec\s g^+]\ra\ra\nonumber\\
&&\la\la {\bf N}_1{\bf N}_3\ra\ra = \left(\la\la\s_0\s_0\ra\ra + 
\la\la\mu_0\mu_0\ra\ra\right)\left(\la\la\mbox{Tr}[\vec\s g]\mbox{Tr}
[\vec\s g]\ra\ra + \la\la\mbox{Tr}[\vec\s g^+]\mbox{Tr}[\vec\s g^+]\ra\ra\right)
\eea
where $\la\la\s_0\s_0\ra\ra =  \la\la\mu_0\mu_0\ra\ra \sim |(v\tau)^2 + x^2|^{1/8}$.

 There are no reasons to think that correlation functions on the chains with the same 
parity  vanish.  Here solitons enter  with Lorentz spin 3/16. The asymptotics 
of the correlation function is given by 
\bea
&&\chi(\tau,x) = \la \re^{\ri\sqrt{2\pi}\Phi_1(\tau,x)}\re^{-\ri\sqrt{2\pi}\Phi_1(0,0)}
\ra \sim \la \re^{\ri\sqrt{2\pi}\Phi_1(\tau,x)}\re^{-\ri\sqrt{2\pi}\Phi_3(0,0)}\ra \sim 
\nonumber\\
&&\rho^{-1/4}\left|\int \rd\theta \re^{3\theta/8}\re^{- m\tau\cosh\theta + 
\ri xm\sinh\theta}\right|^2 = 2\rho^{-1/4}\int\rd\theta K_{3/4}(2m\rho\cosh\theta)
\eea
where $\rho^2 = (v\tau)^2 + x^2$. The Fourier transform is ($s^2 =\omega_n^2 + (vq)^2$)
\bea
&&4\pi\int\rd\rho \rho^{3/4}J_0(s\rho)\int\rd\theta K_{3/4}(2m\rho\cosh\theta) 
\sim \nonumber\\
&&\int\rd\theta (\cosh\theta)^{-7/4}F(5/4,1/2,1;- s^2/4m^2\cosh^2\theta) 
\eea
After the analytic continuation $\ri\omega = \omega + \ri 0$ we get the imaginary part  
(now $s^2 = \omega^2 - (vq)^2 > 4m^2$)
\bea
&&\chi''(\omega,q) \sim s^{-3/4}\int_{(2/s)}^1\frac{\rd x x^{3/4}}
{(1 - x^2)^{3/4}(s^2x^2/4m^2 - 1)^{1/2}}F(1/2,1/2,1/4;1 - x^2)\label{chi}
\eea
Close to the threshold $s = 2m$ we have 
\bea
\chi''(\omega,q) \sim (s^2 - 4m^2)^{-1/4}
\eea
At $s^2 \gg m^2$ Eq.(\ref{chi}) gives the correct asymptotics $\chi'' \sim s^{-1}$ which 
indicates that the two-soliton approximation may give a reasonable description throughout 
the entire range of energies. Recall that for $N =1$ the power was $-1/2$. It is tempting 
to speculate that the threshold singularity  further diminishes with an increase of  $N$.

 The case of four chains introduces some new features. Some of them, as we believe, are 
accidental  and some are generic. The generic feature which persists for higher number of 
chains (see above) is the presence of singlet degrees of freedom. Here they appear in the 
form of the critical Ising model. The accidental features are  the criticality of the latter 
model and the  non-Abelian 
statistics of the massive kinks. These two properties are closely related and are unstable 
with respect to a change in boundary conditions. In the most general setting, as we shall 
see in the next Section, the gap in the Ising model sector will be finite, but much smaller 
than the magnetic gap.   
\medskip

\section{Towards infinite number of chains: a triadic cluster expansion.}
\medskip
 
\begin{figure}[ht]
\begin{center}
\epsfxsize=0.9\textwidth
\epsfbox{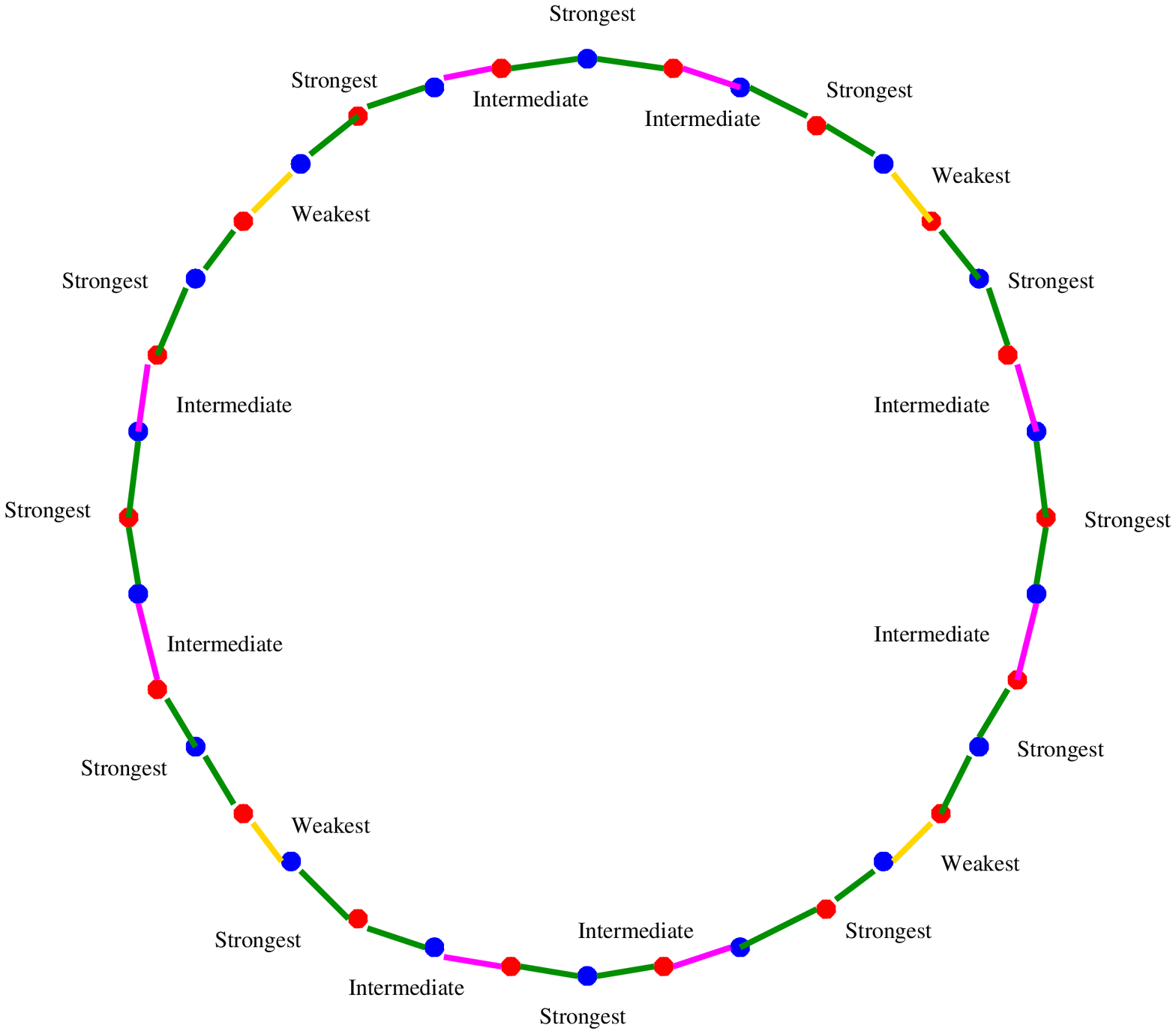}
\end{center}
\caption{ The hierarchical model for 36 chains.  The circles with different color  
denote currents with different chirality. The green bonds are the strongest, the 
magenta ones are of intermediate strength and the yellow ones are the weakest. 
\label{fig:hierarchy}}
\end{figure}

 As the reader probably understands, the problem in question is difficult, naive attempts 
to develop a self-consistent expansion schemes fail (see Section IIIA) and we have to resort 
to some other methods. The exact results for two and four chains discussed in the previous 
Section give a glimpse of the complexity of the problem. Unfortunately, for higher number 
of chains there are no exact solutions. However, a non-perturbative analysis can be imagined, 
though for a somewhat modified model. In this modification  the interactions are set up  
in such a way that the chains are  assembled into clusters  of three (triads), 
nine (enneads), 27 chains {\it etc.}(see Fig.\ref{fig:hierarchy} ).
\newpage

\newpage

 The interaction inside of a cluster of a given size is supposed to be stronger 
than interactions between the clusters of larger sizes. We believe that the study  
of such modified model sheds light on classification of the excitations and establishes 
the structure of the effective action for the collective modes below the soliton mass gap.  

\subsection {The fundamental triad (three chains)}

The solution for two chains with one chirality (let it be the left one) coupled by the
current-current interaction to a chain with opposite chirality was described 
in \cite{jerez} and \cite{azaria} (see also \cite{az}).

The Hamiltonian density is 
\bea
{\cal H} = \left[\frac{\ri v}{2}\bar\chi_0\p_x\bar\chi_0 + 
\frac{\pi v}{2}:\bar{\bf I}\bar{\bf I}:\right] + \frac{2\pi v}{3}:{\bf J J}: + 
\gamma \bar{\bf I}{\bf J}
\eea
where $\bar I^a$ and $J^a$ are right SU$_2$(2) and left SU$_1$(2) currents, respectively. 
The expression in the square brackets describes a sum of two chiral SU$_1$(2) WZNW models. 
Such representation was already discussed in  Section IV.B. The right-moving Majorana 
fermion does not participate in the interaction. In the ultraviolet limit,
the model represents a 
sum of three chiral conformal field theories: the $k =2$ SU(2) WZNW model with the 
right central charge $C_R =3/2$, the right-moving free  Majorana field with $C_R =1/2$ 
and the SU$_1$(2) WZNW model with the left central charge $C_L = 1$. The current-current 
interaction generates a massless RG flow to the infrared 
critical point. The theory in the infrared is represented by two free Majorana 
fermions with opposite chirality $\bar\chi_0$ and $\chi_0$ and the right-moving 
sector of the SU$_1$(2) WZNW model:
\bea
{\cal H}_{IR} = \frac{2\pi v}{3}:{\bf \bar j \bar j}:  + \frac{\ri v}{2}
\left(\bar\chi_0\p_x\bar\chi_0 - \chi_0\p_x\chi_0\right) 
\eea
Under the RG flow the fields transmute in the following way:
\bea
{\bf \bar I} \rightarrow 2{\bf \bar j} + ..., ~~ \ri{\bar\chi_0}\vec{\bar\chi} 
\rightarrow \ri{\bar\chi_0}\chi_0 {\bf \bar j} +... \label{flow}
\eea
where the dots stand for less relevant operators. 
\begin{figure}[ht]
\begin{center}
\epsfxsize=0.45\textwidth
\epsfbox{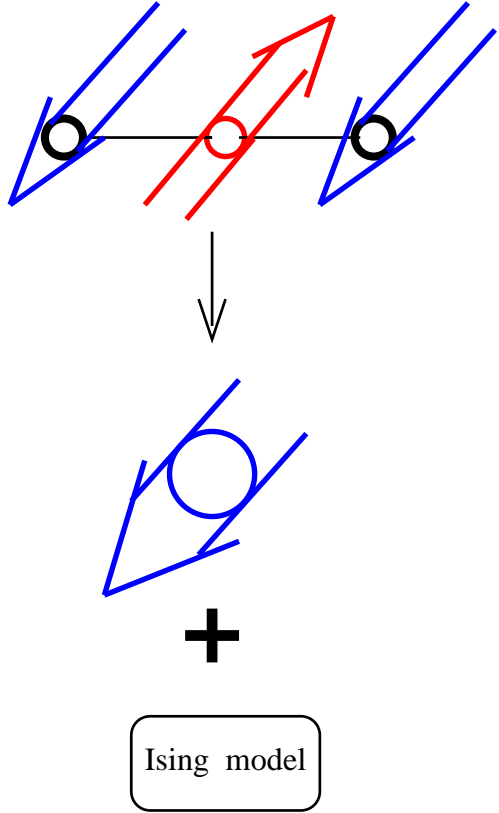}
\end{center}
\caption{ The renormalization scheme for three chains. In the infrared limit the system 
is equivalent to a single chiral C=1 chain and a critical (non-chiral) Ising model. 
The circles with different thickness denote currents with different chirality.
\label{fig:triad}}
\end{figure}

\subsection{The first {\it ennead} (nine).}

 Let us consider the case of nine chains with open boundary conditions, arranging them 
into three triads in such a way that the bare coupling  constant inside each triad 
$\gamma_0$ is greater than the bare coupling   $\gamma_1$ between the triads 
(see Fig. 3). Then we can have a situation when the IR fixed point for a given triad  
is already achieved (the corresponding energy scale is $\Delta \sim 
J_{\parallel}\gamma\exp[- \pi/\gamma]$), well before  
the renormalized coupling for the triad-triad  interaction becomes of the order of 1. 

 According to Eq.(\ref{flow}), the renormalized coupling between any neighbouring   
triads (let us denote them 1 and 2; recall also that they always have different chirality!) 
in the infrared is
\bea
\gamma_1(\Delta){\bf j}_1\bar{\bf j}_2 + \gamma_1(\Delta)
(\ri\bar\chi_0\chi_0)_1(\ri\bar\chi_0\chi_0)_2{\bf j}_1\bar{\bf j}_2 + ... 
\eea
The first term reproduces original  interaction (\ref{model}). Its presence in the 
Hamiltonian also converts  the second term into a   marginal operator:
\bea
&&\gamma_1(\Delta)(\ri\bar\chi_0\chi_0)_1(\ri\bar\chi_0\chi_0)_2{\bf j}_1\bar{\bf j}_2 
= \tilde\gamma_1(\ri\bar\chi_0\chi_0)_1(\ri\bar\chi_0\chi_0)_2 + \gamma_1(\Delta)
(\ri\bar\chi_0\chi_0)_1(\ri\bar\chi_0\chi_0)_2:{\bf j}_1\bar{\bf j}_2: \nonumber\\
&&\tilde\gamma_1 = \gamma_1(\Delta)\la{\bf j}_1\bar{\bf j}_2\ra \label{effinter}
\eea

 Thus by integrating out the high energy degrees of freedom in the {\it ennead}  
of chains one generates  at energies smaller than  $\Delta$ (the new UV cut-off)  
the effective action which contains the original action for a triad of chains  
with a renormalized coupling constant $\gamma_1$ plus the action for three  critical 
Ising models coupled by the products of energy density operators:
\bea
{\cal H}_{Ising} = \sum_{r=1}^3\left[\frac{\ri}{2}(\bar\chi\p_x\bar\chi 
- \chi\p_x\chi)_r\right] + \tilde\gamma_1(\bar\chi\chi)_2[(\bar\chi\chi)_1 
+ (\bar\chi\chi)_3] \label{coupledIs}
\eea
 The Ising  subsystem decouples from the magnetic one. The resulting model is similar 
to the O(3) GN model discussed above (see the discussion around Eq.(\ref{Sm})). 
The only difference is that the symmetry is broken down to U(1). The interaction 
can be written in terms of the currents  
\bea
\tilde\gamma_1[(\bar\chi\chi)_1(\bar\chi\chi)_2 + (\bar\chi\chi)_2(\bar\chi\chi)_3] 
\sim  \gamma_{\perp}[I^x\bar I^x + I^y\bar I^y ] + \gamma_{\parallel}I^z\bar I^z
\eea
which makes the model somewhat similar to the anisotropic Thirring model ( $I^a$ are 
SU$_2$(2) currents made of $\chi_r$'s and the coupling constants $\gamma_{\perp} 
\sim \tilde\gamma_1$ and $\gamma_{\parallel} =0$ on the bare level). We believe 
(the details will be given in a separate publication), that this particular type 
of anisotropy vanishes in the strong coupling regime in the same way as it does 
in the C-sector of anisotropic Thirring model \cite{JNW},\cite{alts}. In other 
words in the strong coupling regime the model (\ref{coupledIs}) is equivalent 
to the O(3) GN model. 

  Thus on each step of the triadic real space RG we generate the O(3) GN models. 
Their number is equal to the number of clusters on the given level. For example, 
if the initial number of chains is 3$^N$ and the first GN model appears for a cluster 
of 9 chains, the number of copies of  GN models on the first level is 3$^{N -2}$. 
Then  the next levels contain 3$^{N -3}$, 3$^{N - 4}$ {\it etc.} copies such that 
the total number of such models in both parity sectors is 3$^{N -1}$. Taking into 
account that each of them has 2-fold degenerate ground state, we get the ground 
state degeneracy 2$^{{\cal N}/3}$, where ${\cal N} = 3^N$ is the total number 
of chains. The resulting ground state entropy is three times smaller than the one 
which was obtained for the uniform case. 

\begin{figure}[ht]
\begin{center}
\epsfxsize=0.45\textwidth
\epsfbox{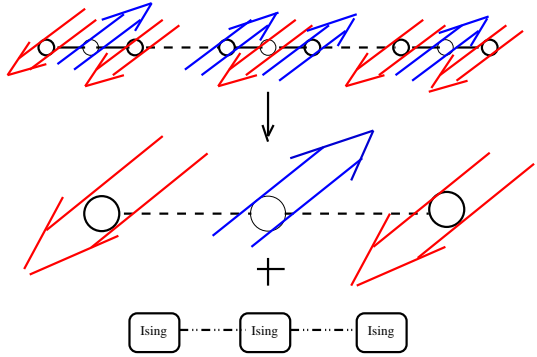}
\end{center}
\caption{ A schematic picture of renormalization of nine chains. 
\label{fig:ennead}}
\end{figure}

\begin{figure}[ht]
\begin{center}
\epsfxsize=1.0\textwidth
\epsfbox{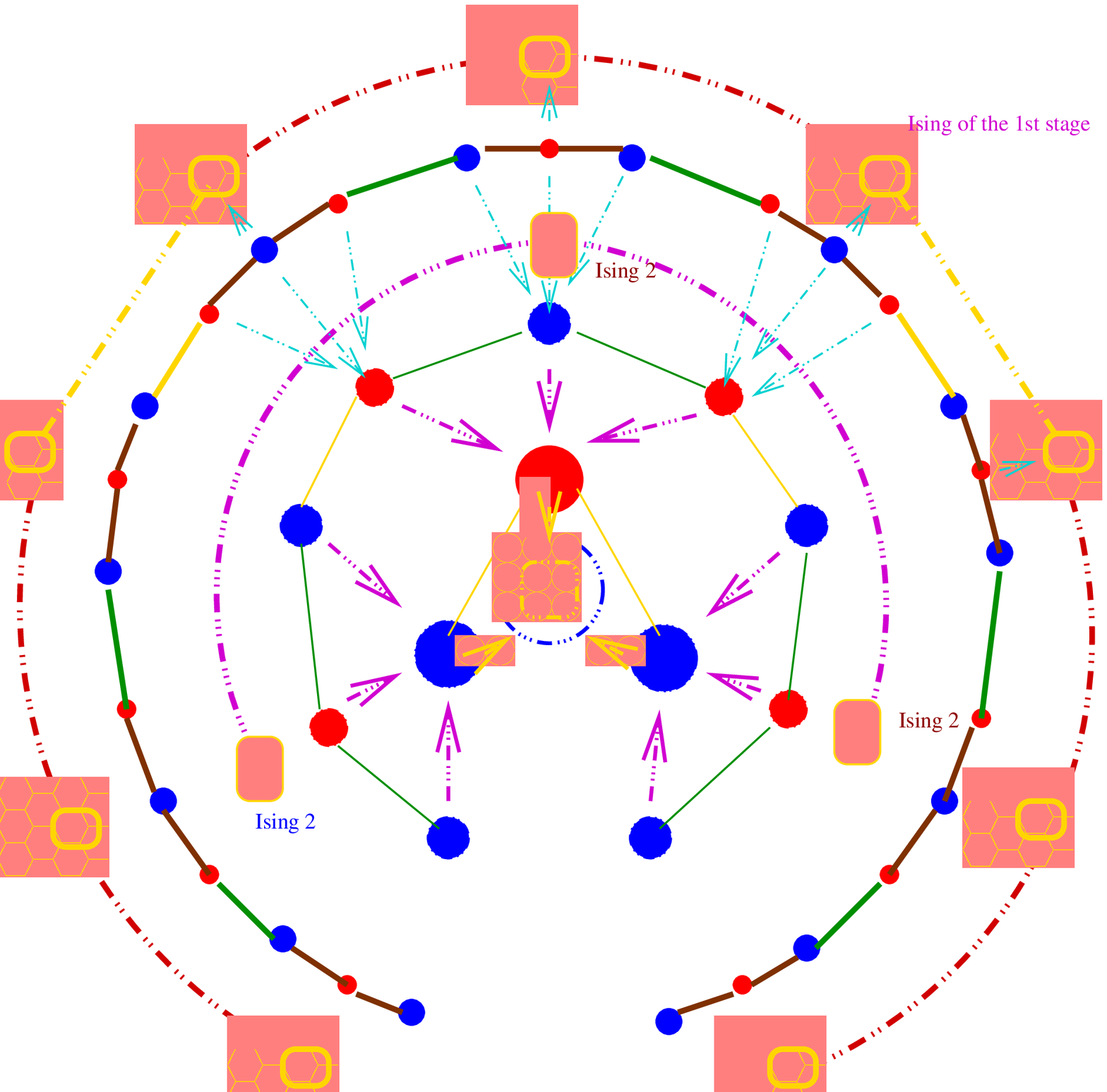}
\end{center}
\caption{The hierarhy of interactions and the renormalization process for 27 chains. 
The strongest bonds are brown, the intermediate are gree and the weakest ones 
are yellow. 
The Ising variables are shown as pink rectangles. The red and blue circles of 
various sizes 
depict spin currents of different chirality. Their size increases with every step of 
the RG process. \label{fig:C
}}
\end{figure}

 Now we can discuss the thermodynamics. In our cluster expansion we have the following 
energy scales. First, there is a sequence of crossover scales $\Delta_0>  
\Delta_1, ... > \Delta_{N -1}$ (magnetic gaps) corresponding to crossovers inside of
 each cluster. Then there are energy gaps of the interacting Ising models $M_2, M_3, ...$ 
which are formed by clusters of 9, 27, 81, {\it etc.} chains. The first scale in 
this sequence is of the order of  
\bea
M_{2} \sim \Delta_1\exp[- \pi/\tilde\gamma_1]
\eea
Since $\tilde\gamma_1 \sim \gamma_1^2$ we deem these energy gaps to be much smaller 
than the magnetic gaps. We conjecture  that this difference survives even in the limit 
when all interactions become equal.

 All these crossovers affect the temperature behavior of the specific heat. The first 
crossover occurs at temperature $T \sim \Delta_0$ which is the crossover temperature of 
a fundamental triad. Above this temperature we have a bunch of non-interacting chiral 
Heisenberg chains each having the  heat capacity linear in $T$:
\bea
C(T > \Delta_0) = S{\cal C}, ~~ {\cal C} = \frac{\pi T}{6v}  
\eea
where $S = 3^NL_{\parallel}$ is the total area occupied by the system.
At  $\Delta_0 > T > \Delta_1$ we effectively have $3^{N -1}$ critical Ising chains with 
central charge $C = 1/2$ and the same amount of non-interacting chiral Heisenberg chains. 
As a result the slope of the {\it specific heat} drops by the factor of 2. When the 
entire sequence of magnetic crossovers is passed, that is at $\Delta_{N -1} > T > M_2$ 
the specific heat is given only by the Ising models:
\bea
{\cal C} = \frac{\pi T}{6v}(\frac{1}{3} + \frac{1}{9} + \frac{1}{27} + ...) = 
\frac{\pi T}{12v}
\eea
that is remains the same as after the first crossover. Thus we can say that in the 
thermodynamic limit the singlets ocuppy a half of the original Hilbert space.
\begin{figure}[ht]
\begin{center}
\epsfxsize=0.45\textwidth
\epsfbox{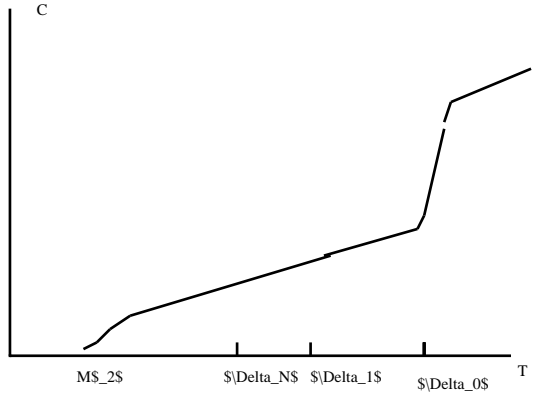}
\end{center}
\caption{ A schematic picture of the temperature dependence of the specific heat 
\label{fig:C1
}}
\end{figure} 

 When the temperature falls below the first singlet gap $M_2$, one may say to the 
first approximation that the Ising modes with the largests gap cease to contribute. 
After a certain crossover the linear slope in the specific heat drops by the factor of 9. 
The specific heat remains linear in $T$ until $M_3$ is reached where the slope falls 
by another factor of 3 {\it etc.} In our scheme where the distribution of coupling 
constants between the clusters of different size is rather arbitrary, it makes little 
sense to do more detailed calculations. The only thing we can say is that $C \sim T$ 
above certain temperature and than experiences a fast decrease. In our cluster expansion 
there are two areas with linear specific heat characterised by the slopes differing by 
the factor of 2. This due to the fact that within this approach  the gap for magnetic 
excitations $\Delta$ is different from the gaps for singlet excitations. Whether this 
feature will survive in the limit of uniformly coupled chains is open for debate.  
 
\section{Some leftovers}

 It would make a lot of sense to study the models described in this paper numerically. 
Since numerical calculations will probably be performed for systems with restricted 
number of chains, we decided to describe certain tractable cases in more detail.

\subsection{An approximate solution for $N =3$ (six chains).}

 In this case the magnetic subsystem is equivalent to $N =1$ (that is to the SU(2) 
Thirring model) and the Ising subsystem contains two Ising models with the 
$\epsilon_1\epsilon_2$ interaction. The latter is equivalent to the spinless 
Thirring model with a truly marginal interaction. Thus the  Ising sector is critical 
and is described by the C =1 Gaussian model.

\subsection{An approximate solution for $N =6$ (twelve chains).}

 Applying the procedure outlined above to 
the system of twelve chains with periodic boundary conditions (see Fig. 3), we obtain 
the $N =2$ model in the magnetic sector plus model (\ref{coupledIs}) with four chains.  
\begin{figure}[ht]
\begin{center}
\epsfxsize=0.65\textwidth
\epsfbox{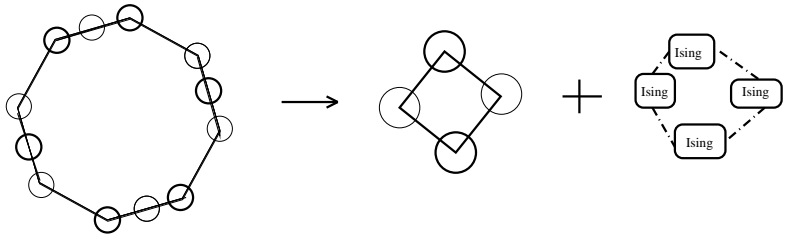}
\end{center}
\caption{ A schematic picture of twelve chains. The fat lines correspond to larger exchange 
interactions. The circles of different size denote chains with different chirality. 
\label{fig:twelve
}}
\end{figure}
 
 Thus the excitation spectrum includes GN O(3) soliton mode, one gapless Majorana fermion 
and massive excitations from the Ising sector.

\section{Single-electron Green's function for the model of stripes and the spinon propagator 
for $N = \infty$.}

 Now we shall try to use the wisdom accumulated in the exact solution of $N =2$ case 
to obtain 
results for the case of infinite number of chains. Though our ultimate goal is to 
calculate 
the correlation function of staggered magnetizations, the path to it lies through  
the single-electron Green's function,  which exists only for the model of stripes 
(Fig. 1b). 
The reason for this will become clear in the process of calculation.

Let conduction electrons belong to even chains.  Then the  electron creation  operator is 
\bea
R_{\s,2n} = \re^{\ri\sqrt{2\pi}\varphi_c(2n)}\re^{\s\ri\sqrt{2\pi}\varphi(2n)}, 
~~L_{\s,2n} = \re^{-\ri\sqrt{2\pi}\bar\varphi_c(2n)}\re^{-\s\ri\sqrt{2\pi}\bar\varphi(2n)}
\label{factorization}
\eea
where $v_c$ is the Fermi velocity. Though, in general, $v_c$ may  be quite different from 
the spin velocity, we shall not consider this possibility. With $v_c = v_s$ the model is 
(1+1)-dimensional Lorentz invariant. 
The charge fields $\varphi_c, \bar\varphi_c$ are free Gaussian fields and hence there 
are no correlations between the fields belonging to different chains. 
\bea
G_{RR}(\tau,x) = \frac{1}{(\tau v_c + \ri x)^{1/2}}G_{2n,2n}(\tau,x), 
~~G_{LL}(\tau,x) = \frac{1}{(\tau v_c - \ri x)^{1/2}}G_{2n,2n}(\tau,-x) \label{singleP}
\eea
where function $G$ was defined in Eq.(\ref{factor1}) and can be called the spinon 
Green's function.  
Since the electron Green's  function must be a single-valued function of $x$ at 
$\tau \rightarrow 0$ and a single-valued function of $\tau$ at $x =0$, we conclude 
that the spinon function $G$ must be a double-valued function to compensate for the 
double-valuedness of $(\tau v_c \pm \ri x)^{-1/2}$. This suggests that the spinons 
are semions, as it was suggested by Laughlin \cite{Laughlin}.

Now we shall use these facts to determine the asymptotics of the single-fermion 
Green's function in a similar way as it was done for the (1+1)-dimensional Thirring 
model \cite{singleP},\cite{LukZam}. Namely, we shall combine these arguments with 
the Lehmann expansion for the Green's function. In this expansion we shall take 
into account only terms with an emission of a single massive soliton. The spectrum 
of this soliton is 
\bea
E(k,k_{\perp}) = \sqrt{(vk)^2 + M^2(k_{\perp})}
\eea
where $M(k_{\perp}) \sim \exp[-\pi v/\gamma]$ is as yet unknown function (the soliton gap). 
We believe that this gap always remains finite for any momentum. A convenient 
parametrization of the energy and the momentum component along the chain direction is 
\bea
E = M(k_{\perp})\cosh\theta, ~~ vk_x = M(k_{\perp})\sinh\theta
\eea
In these notations a Lorentz rotation on an 'angle' $\gamma$ corresponds to the shift 
of rapidity $\theta \rightarrow \theta + \gamma$. Since an operator with Lorentz spin $S$ 
transforms under such  Lorentz transformation as 
\[
A_S \rightarrow \re^{\gamma S}A_S
\]

We generalize for $N = \infty$ the formulae obtained for $N=1$ and $N =2$ case 
treating the vanishing of $D$-function is the $N =2$ as an artefact of periodic 
boundary conditions. 
\bea
&&G  \sim (v\tau -\ri x)^{-1/2}Z(k_{\perp})\exp[- M(k_{\perp})\sqrt{\tau^2 + 
(x/v)^2}] \nonumber\\
&&D \sim f(k_{\perp})M^{1/2}(k_{\perp})K_0[M(k_{\perp})
\sqrt{\tau^2 + (x/v)^2}] 
\eea
where $Z,f$ are yet unknown functions. The $f$-function may vanish at some point in 
momentum space, as suggested by the example of four chains.
Notice that both $k_{\perp}$-dependence of $G$ and the very existence of $D$ are 
effects exponential in $1/\gamma$. These equations 
 dictate the following asymptotic form for 
 the correlation function of the staggered magnetizations:
\bea
&&\chi_1(\tau,x;q_{\perp}) = \rho^{-1}\int\frac{\rd k_{\perp}}{(2\pi)}Z(k_{\perp})
Z(q_{\perp} + k_{\perp})\exp\{- [M(k_{\perp}) + M(q_{\perp} + k_{\perp})]\rho\}\\
&&\chi_2(\tau,x;q_{\perp}) = \int\frac{\rd k_{\perp}}{(2\pi)}f(k_{\perp})f(k_{\perp} 
+ q_{\perp})[M(k_{\perp})M(q_{\perp}]^{1/2}K_0[M(k_{\perp})\rho]K_0
[M(k_{\perp} + q_{\perp})\rho]\nonumber
\eea
where $\rho^2 = \tau^2 + (x/v)^2$ and $\chi_1$. is a correlation function between 
the chains with the same and $\chi_2$ with different parities. 
 
\section{Conclusions}

 Let us make a summary of our results. 

\begin{itemize}
\item
We have  found  models of magnets with a short range Heisenberg exchange which have 
neither magnetic  nor spin-Peierls  order.
\item
We have given a formal prove that these models possess 
excitations with spin 1/2.
\item
We established the existence of $T =0$ critical point and a massive ground state 
degeneracy. The ground state entropy is proportional to the number of chains. 
\item
 We have also established that the  low-energy  Hamiltonian  separates into two 
weakly interacting parts describing sectors 
with  different parity.
\item
We established that the correlation functions of staggered magnetizations can be 
written as real space products of correlation functions of non-local 
operators belonging to the sectors with different parity (see Eqs.(\ref{factor1},
\ref{factor2})). 
\item
We have solved the problem exactly for the case of two and four coupled chains 
(the latter one with periodic boundary conditions). The results are consistent 
with  the statements made for the case of infinite number of chains. 
The spectrum contains a massive amount of singlet excitations. In the language 
of gauge theory these excitations describe dynamics of Wilson loops. 

\end{itemize}

 A natural question is whether the models considered in our work can be realized.  
We believe that the the answer is positive. The model of stripes may well be 
relevant in a strongly underdoped regime  of copper oxides. 

 We have already mentioned in Introduction, that there are other models of 
fractionalization, 
such as suggested in \cite{fisher},\cite{nayack} together with the corresponding 
dimer models 
mentioned earlier. Though these works use Hamiltonians not based on any 
microscopic electron 
models, one might hope that they capture  some  features of the solution. 
In particular, it 
would be interesting to study in detail how singlet  excitations, which in 
 our model are  
associated with the O(3) GN models emerging on each triad of chains, are 
related to Z$_2$ vortices 
(visons) introduced in \cite{fisher},\cite{read}(see also  the recent work \cite{iosele}, 
which essentially clarifies the concept of vison). We leave this question, as many others, 
to the future research.

\section{ Acknowledgements}
We are grateful to  Alexander Abanov, Boris Altshuler, Claudio Castellani, 
Andrey Chubukov, 
Fabian Essler, Michele Fabrizio, Vladimir Fateev, Eduardo Fradkin, Lev Ioffe, 
Dmitrii Khveshchenko, Vladimir Kravtsov, Feodor Smirnov, Peter Fulde, Maurice Rice, 
Nic Shannon,  
Oleg Starykh  and Arkadii Vainshtein for valuable discussions and interest to 
the work. A. M. T. 
is also grateful to Steve Kivelson for sending him his work before publication.  
A. A. N. and A. M. T. are grateful to Brookhaven National Laboratory 
and  the Abdus Salam ICTP, respectively, for their hospitality. 
We  also acknowledge the support from 
US DOE under contract number DE-AC02 -98 CH 10886.

\section{Appendix A. Gauge theory formulation.}
 The  
relationship with the gauge theories can be inferred from 
the fact that the level $k=1$ SU(2) Kac-Moody currents can be 
identified with
(iso)spin currents of  massless Dirac fermions:
\bea
J^a = \frac{1}{2}R^+_{\alpha}\s^a_{\alpha\beta}R_{\beta}, ~~ \bar J^a = 
\frac{1}{2}L^+_{\alpha}\s^a_{\alpha\beta}L_{\beta},
\eea
$\s^a$ being Pauli matrices. 
The Hamiltonian (\ref{WZNW}),(\ref{model}) becomes
\bea
H = \sum_n\int\rd x\left[\ri v\left(- R^+_{\alpha}\p_xR_{\alpha} + 
L^+_{\alpha}\p_xL_{\alpha}\right)_n + \frac{\gamma}{8}\sum_{\mu}(R^+{\vec\s}R + 
L^+{\vec\s}L)_n \cdot (R^+{\vec\s}R + L^+{\vec\s}L)_{n + \mu}\right]\label{fermions}
\eea
As gauge theories of the RVB state, this theory possesses redundant charge degrees of 
freedom which do not participate in the interactions.  Since every approximation 
violates this subtle property, dealing with the charge sector 
would become the same  awkward 
problem as it is in  the standard approach  to the RVB  gauge theory, 
once one decides 
to 
adopt this fermionic representation.  Let us, however, follow the well-trodden
 path for a while just to 
make sure that the model we are discussing does fall in the category of 
RVB liquids.
To this end, we use the identity
\[
\s^a_1\s^a_2 = 2P_{12} - 1,
\]
where $P_{12}$ is the permutation operator, and apply the Hubbard-Stratonovich 
transformation to rewrite the interaction term in Eq.(\ref{fermions}) as
\bea
\sum_{\mu,n}\int \rd x \left[\frac{|\Delta_{\mu +n,n}(x)|^2}{2\gamma} + 
\left(\Delta_{n,n + \mu}R^+_{\alpha,n}L_{\alpha,n + \mu} + H.c.\right)\right] + ...  
\eea
where the dots stand for the terms we deem irrelevant. The procedure essentially 
coincides with the conventional decoupling scheme 
in the RVB  approach. The fact that such 
decoupling here is done only in one lattice direction is not important provided 
one can justify that fluctuations  of $|\Delta|$ may be  neglected. As we shall see, 
excitations 
associated with breaking of singlets carry the largest spectral gap.  This 
justifies the assumption about small fluctuations of $|\Delta|$.  Once the 
amplitude $|\Delta| \sim J_{\parallel}\exp(- \pi v/\gamma)$ 
 is frozen, one is  left with the compact 
U(1) lattice 
gauge theory in the strong coupling limit (indeed, the gauge field has no 
bare  
$F^2_{\mu\nu}$-term which corresponds to  infinite bare charge.). The vector 
potential 
$A_y$ is represented by the phase of $\Delta_{n,n + \mu}$. Since by omitting 
the term with 
$R^+_{\alpha}R_{\alpha}L^+_{\alpha}L_{\alpha}$ we violate the decoupling of 
the charge 
degrees of freedom, we have to enforce the constraint on the absence of charge fluctuations 
by  introducing   the time component $A_0$ of the gauge field. In the mean field 
approximation (that is, when  fluctuations of the gauge fields are neglected) we  
obtain the  spectrum of the $\pi$-flux state \cite{affleck}: 
\bea
E(p) = \sqrt{(vk_{\parallel})^2 + 4|\Delta|^2\cos^2k_{\perp}}\label{MF}
\eea
which has two Dirac-like conical singularities in the Brillouin zone. From what 
we have done in this paper it appears very dubious that this result will survive
 inclusion of fluctuations.

\section{Appendix B. The structure of strong coupling regime and the 
ground state degeneracy.}

Let us assume that the system is two-dimensional and the number of chains
is even, $2N$.
We will  drop the Lorentz-noninvariant part, ${\cal L}^{int}_n$, in which case
one has a decomposition
${\cal L}_n = {\cal L}^{+}_n + {\cal L}^{-}_n$.

Let us first consider $\sum_n {\cal L}^{+}_n$.
In the strong-coupling ground state, the following combinations of
the chiral fields are locked:
\be
\varphi_{2n} + \bar\varphi_{2n+\mu} = \sqrt{\frac{\pi}{8}}
\left[1 + 2 m_{2n}^{(\mu)}  \right], ~~~\mu = \pm 1 ,
~~~ m_{2n}^{(\mu)} = 0, \pm 1, \pm 2, ...
\label{locked}
\ee
Assuming that periodic boundary conditions are imposed in the transverse
direction, the set of equations (\ref{locked}) can be rewritten as 
\bea
\varphi_{0} + \bar\varphi_{2N-1} &=& \sqrt{\frac{\pi}{8}}
\left[ 1 + 2 m_{0}^{(-)} \right], \nn\\
 \varphi_{0} + \bar\varphi_{1} &=& \sqrt{\frac{\pi}{8}}
\left[ 1 + 2 m_{0}^{(+)} \right], \nn\\
 \varphi_{2} + \bar\varphi_{1} &=& \sqrt{\frac{\pi}{8}}
\left[ 1 + 2 m_{2}^{(-)} \right], \nn\\
 \varphi_{2} + \bar\varphi_{3} &=& \sqrt{\frac{\pi}{8}}
\left[ 1 + 2 m_{2}^{(+)} \right], \nn\\
\cdot ~~\cdot ~~\cdot~~ &&  ~\cdot ~\cdot ~~\cdot ~~\cdot ~~\cdot  \nn\\
 \varphi_{2N-2} + \bar\varphi_{2N-3} &=& \sqrt{\frac{\pi}{8}}
\left[ 1 + 2 m_{2N-2}^{(-)} \right], \nn\\
 \varphi_{2N-2} + \bar\varphi_{2N-1} &=& \sqrt{\frac{\pi}{8}}
\left[ 1 + 2 m_{2N-2}^{(+)} \right].
\label{locking-eqs}
\eea
Considering pairs of neighboring equations in (\ref{locking-eqs})
 one finds that the integers
$m_{2n}^{(\mu)}$ satisfy the condition:
\be
\sum_{k=0}^{N-1} m^{(+)}_{2k} = \sum_{k=0}^{N-1} m^{(-)}_{2k}.
\label{integ-condition}
\ee
With this constraint, the  number of independent fields, 
locked in the ground state, is $2N - 1$.
To select these fields
we first
define $N$ scalar fields 
\be
\chi_n = \varphi_{2n} + \bar\varphi_{2n+1}, ~~n = 0,1,2, ... , N-1
\label{chi-n}
\ee
together with their $N$ dual counterparts
\be
\vartheta_n = \varphi_{2n} - \bar\varphi_{2n+1}.
\ee
According to (\ref{locked}), the fields $\chi_n$ are locked at
\be
\left[ \chi_n \right]_{\rm vac} =
\sqrt{\frac{\pi}{8}} \left[1 + 2 m^{(+)}_{2n}  \right].
\label{chi-locked}
\ee
On the other hand,
there are $N$ more combinations,
$
\varphi_{2n} + \bar\varphi_{2n-1},
$
that get frozen, again according to (\ref{locked}). We can express them in terms
of $\chi_n$ and $\vartheta_n$:
\be
\varphi_{2n} + \bar\varphi_{2n-1} =
\frac{1}{2} \left( \chi_n +  \vartheta_n +  \chi_{n-1} -  \vartheta_{n-1} \right)
= \sqrt{\frac{\pi}{8}}\left[1 + 2 m^{(-)}_{2n}  \right].
\label{rel-aux}
\ee
Thus $N$ relative dual fields
\be
\theta^{(-)}_n = \frac{1}{\sqrt{2}} \left(\vartheta_{n} - \vartheta_{n+1}  \right),
~~~n = 0,1,2, ... , N - 1
\ee
are also locked:
\be
\left[ \theta^{(-)}_n \right]_{\rm vac}
= \frac{\sqrt{\pi}}{2} \left[ m^{(+)}_{2n} + m^{(+)}_{2n + 2}
- 2 m^{(-)}_{2n + 2} \right]. \label{theta-locked}
\ee
Notice that transverse periodic boundary conditions imply that
\[
\sum_n \theta^{(-)}_n = 0,
\]
(which is actually the same condition as Eq.(\ref{integ-condition}))
and so the number of independent relative fields is $N - 1$.
We choose them to be
\be
\theta^{(-)}_{0}, ~\theta^{(-)}_{1}, ~...~ \theta^{(-)}_{N-2}
\label{N-1:thetas}
\ee
The total field
\be
\theta^* = \frac{1}{\sqrt{N}} \sum_{n=0} ^{N-1} \vartheta_n
\label{theta-total}
\ee
remains unlocked and, hence, disordered.
\medskip

Thus the $N$ fields $\chi_n$, Eq.(\ref{chi-n}), 
are locked while $N$ their dual counterparts $\vartheta_n$ are {\sl not};
only thier $N-1$ combinations (\ref{N-1:thetas}) are. 
The frozen values of these $2N - 1$ independent fields 
characterize the vacuum state of $\sum_n {\cal L}^{+} _n$.
\medskip

The same can be done for $\sum_n {\cal L}^{-} _n$: we introduce
\bea
\psi_n &=& \bar\varphi_{2n} + \varphi_{2n+1},  \label{Psi-s} \\
\omega_n &=& - \bar\varphi_{2n} + \varphi_{2n+1} \nn
\eea
and select $N$ fields $\psi_n $ and $N-1$
relative duals
\be
\omega^{(-)}_{0}, ~\omega^{(-)}_{1}, ~...~, ~\omega^{(-)}_{N-2},
\label{N-1:omegas}
\ee
where
\[
\omega^{(-)}_n = \frac{\omega_n - \omega_{n+1}}{\sqrt{2}}.
\]
The total field
\[
\omega^* = \frac{1}{\sqrt{N}} \sum_{n=0}^{N-1} \omega_n
\]
remains disordered.
The $\psi_{n}-$fields are frozen at
\be
\left[ \psi_n \right]_{\rm vac} = \sqrt{\frac{\pi}{8}}
\left[ 1 + 2 m^{(-)}_{2n+1} \right] \label{psi-locked}
\ee
while $\omega^{(-)}_n$ -- at
\be
\left[\omega^{(-)}_n \right]_{\rm vac} = \frac{\sqrt{\pi}}{2} 
\left[2 m^{(+)}_{2n+1} - m^{(-)}_{2n+1} -  m^{(-)}_{2n+3}  \right].
\label{omega-rel-locked}
\ee
\bigskip

We can now express the physical fields of individual chains
in terms of the locked fields and two extra fields that remain disordered.
We have:
\bea
\Phi_{2n} &=& \frac{1}{2} \left( \chi_n + \psi_n + \vartheta_n - \omega_n \right), \nonumber\\
\Theta_{2n} &=& 
\frac{1}{2} \left( \chi_n - \psi_n + \vartheta_n + \omega_n \right), \nonumber\\
\Phi_{2n+1} &=& \frac{1}{2} \left( \chi_n + \psi_n - \vartheta_n + \omega_n \right), 
\nonumber\\
\Theta_{2n+1} &=& \frac{1}{2} \left( -\chi_n + \psi_n + \vartheta_n + \omega_n \right)
\label{phys-fields}
\eea
The elements of the Wess-Zumino matrix field $\hat{g}_n (x)$ contain
exponents $e^{\pm i \sqrt{2\pi} \Phi_n}$ and
$e^{\pm i \sqrt{2\pi} \Theta_n}$.
According to  Eqs. (\ref{phys-fields}),
in the strong coupling phase these exponents are proportional to
\[
\exp \left(\ri \sqrt{\frac{2\pi}{N}} \theta^*\right) ~~{\rm and} ~~ 
\exp \left(\ri \sqrt{\frac{2\pi}{N}} \omega^*\right), 
\]
respectively,
and thus will have vanishing expectation values. 
This effect, however, 
disappears in the thermodynamic limit. 
\bigskip

{\bf Transverse dimerization and degeneracy of the ground state}

Spontaneous transverse dimerization has been identified 
for the case of two chains in
Ref.\cite{allen}.
In the continuum limit, the corresponding order parameter is given by
a simple expression: $\bN_n (x) \cdot \bN_{n+1} (x)$. {\sf For the 2N-chain model}
\bea
\bN_{2n} \cdot \bN_{2n+1}  
&\sim&  \cos \sqrt{2\pi} (\Theta_{2n} - \Theta_{2n+1}) \nn\\
&+& \frac{1}{2} \left[ \cos \sqrt{2\pi} (\Phi_{2n} - \Phi_{2n+1})
- \cos \sqrt{2\pi} (\Phi_{2n} + \Phi_{2n+1}) \right] \nn\\
&=& \cos \sqrt{2\pi} (\chi_n - \psi_n)
+ \frac{1}{2} \left[ \cos \sqrt{2\pi} (\vartheta_n - \omega_n)
- \cos \sqrt{2\pi}(\chi_n + \psi_n) \right]. \nn
\eea
Using Eqs.(\ref{chi-locked}) and 
(\ref{psi-locked}) when averaging over the ground state, one finds that
\bea
\eta_{2n} &\equiv& \la \bN_{2n} \cdot \bN_{2n+1} \ra
\sim \la  \cos \sqrt{2\pi} (\chi_n - \psi_n) \ra -
\frac{1}{2} \la  \cos \sqrt{2\pi} (\chi_n + \psi_n) \ra \nn\\
&\propto& \frac{3}{2} (-1)^{m_{2n}^{+} - ~m_{2n+1}^{-}}
\label{trans-dim-op-even}
\eea
We observe the existence of two, doubly degenerate values
of this local order parameter; its sign depends on the parity of
the integer $m_{2n}^{+} - m_{2n+1}^{-}$.

Similarly
\bea
&&\bN_{2n+1} \cdot \bN_{2n+2} 
 \sim
\cos \sqrt{2\pi} (\Theta_{2n+1} - \Theta_{2n+2}) \nn\\
&&~~~~~~~~~~
+ \frac{1}{2} \left[ \cos \sqrt{2\pi} (\Phi_{2n + 1} - \Phi_{2n+2})
- \cos \sqrt{2\pi} (\Phi_{2n+1} + \Phi_{2n+2}) \right] \nn\\
&&~~\sim \cos \sqrt{\frac{\pi}{2}} \left[
\left(\psi_n + \psi_{n+1}  \right) - \left(\chi_n + \chi_{n+1}  \right)
+ \sqrt{2} \left( \theta^- _n + \omega^- _n \right) 
\right] \nn\\
&& ~~+ \frac{1}{2} \cos  \sqrt{\frac{\pi}{2}} \left[
\left(\psi_n - \psi_{n+1}  \right) + \left(\chi_n - \chi_{n+1}  \right)
+ \left( \omega_n + \omega_{n+1} \right) - \left( \theta_n + \theta_{n+1} \right)\right]
\nn\\
&& ~~ - \frac{1}{2} \cos  \sqrt{\frac{\pi}{2}} \left[
\left(\psi_n + \psi_{n+1}  \right) + \left(\chi_n + \chi_{n+1}  \right)
- \sqrt{2} \left(\theta^- _n - \omega^- _n \right)  
\right]
\label{interm}
\eea

The first and third terms in the r.h.s. of (\ref{interm})
have nonzero expectation values. Using the locked values of the corresponding fields
given above, we find that arguments of these two cosines are
$\pi \left[  m^{(+)}_{2n+1} - m^{(-)}_{2n+2}\right]$ and 
$\pi \left[  m^{(+)}_{2n+1} + m^{(-)}_{2n+2} + 1\right]$,
respectively.
Therefore
\be
\eta_{2n+1} = \la \bN_{2n+1} \cdot \bN_{2n+2}  \ra
\propto \frac{3}{2} (-1)^{m^{(+)}_{2n+1} - m^{(-)}_{2n+2}}
\label{trans-dim-op-odd}
\ee

Inspecting (\ref{trans-dim-op-even}) and
(\ref{trans-dim-op-odd}) we observe that 
the two signs of local order parameters $\eta_m$ 
reflect a spontaneously broken $Z_{2}$ symmetry.
This is the symmetry related to independent translations by one lattice
spacing along the chains.
Notive, however, that for different pairs of chains
the signs of the order parameters are {\sl uncorrelated}. 
The only condition imposed on the $\eta$'s is
\be
\prod_{n=0}^{2N} \eta_n \propto (-1)^{Q} = 1
\label{constr}
\ee
where the integer
\[
Q = \sum_{\mu = \pm} \sum_{m=0}^{N-1} \mu \left[m^{\mu}_{2m} - m^{\mu}_{2m+1}  \right]
\]
vanishes according to relations (\ref{integ-condition}) in the
(+) sector and their counterparts in the (--) sector.
Eq.(\ref{constr})
is not a restrictive condition; it simply says that the numbers
of positive and negative $\eta$'s should be even. 
Thus the ground state of the system exhibits a huge degree of degeneracy.
Taking
the constraint (\ref{constr}) into account,
the number of the degenerate ground-state configurations in a system of 2N chains is
estimated as
\[
\frac{1}{2}\sum_{n=0}^{2N} C^{n}_{2N} = \frac{1}{2}\sum_{n=0}^{2N} 
\frac{(2N)!}{n! (2N - n)!}
= 2^{2N -1}.
\]

\end{document}